\documentclass[10pt, conference, letterpaper]{IEEEtran}
\IEEEoverridecommandlockouts

\usepackage{tikz}
\usepackage{amsmath,amsfonts,mathrsfs,amssymb}
\usepackage{graphicx,subfig}
\usepackage[normalem]{ulem}
\usepackage[linesnumbered,lined,boxed, commentsnumbered, ruled]{algorithm2e}
\usepackage{algorithmic}
\usepackage{booktabs}
\usepackage{multirow}
\usepackage{amsopn}
\usepackage{etoolbox}
\usepackage{xcolor}
\usepackage{url}

\usepackage{xspace}
\usepackage{indentfirst}
\usepackage{microtype}
\usepackage{color}
\usepackage{listings}
\usepackage{enumitem}
\usepackage{diagbox}
\usepackage{colortbl}
\usepackage{pifont}
\usepackage[misc]{ifsym}

\newcommand{\name}{{\sc MegaReduce}\xspace}

\newcommand{\etc}{\emph{etc.}\xspace}
\newcommand{\ie}{\emph{i.e.,}\xspace}
\newcommand{\eg}{\emph{e.g.,}\xspace}
\newcommand{\etal}{\emph{et al.}\xspace}

\newcommand{\todo}[1]{\textbf{\textcolor{red}{(TODO:)#1}}}

\begin{document}

\title{\emph{Your Mega-Constellations Can Be Slim:} \\
	A Cost-Effective Approach for Constructing Survivable and Performant LEO Satellite Networks}


\author{
	\IEEEauthorblockN{Zeqi Lai$^\ddag$$^\ast$, Yibo Wang$^\ddag$, Hewu Li$^\ddag$$^\ast$, Qian Wu$^\ddag$$^\ast$, Qi Zhang$^\ast$, Yunan Hou$^\ddag$, Jun Liu$^\ddag$$^\ast$, Yuanjie Li$^\ddag$$^\ast$}
	\IEEEauthorblockA{
		\textit{$^\ddag$Institute for Network Sciences and Cyberspace}, Tsinghua University,
			$^\ast$Zhongguancun Laboratory
		%
	}
}

\maketitle

\begin{abstract}





Recently we have witnessed the active deployment of mega-constellations with hundreds to thousands of low-earth orbit~(LEO) satellites, targeting at constructing LEO satellite networks~(LSN) to provide ubiquitous Internet services globally. However, while the massive deployment of LEO satellites can improve the network survivability and performance of an LSN, it also involves additional \emph{sustainable challenges} such as higher deployment cost, risk of satellite conjunction and space debris.

In this paper, we investigate an important research problem facing the upcoming satellite Internet: \emph{from a network perspective, how many satellites exactly do we need to construct a survivable and performant LSN?} To answer this question, we first formulate the survivable and performant LSN design~(SPLD) problem, which aims to find the minimum number of needed satellites to construct an LSN that can provide sufficient amount of redundant paths, required link capacity and acceptable latency for traffic carried by the LSN. Second, to efficiently solve the tricky SPLD problem, we propose \name, a requirement-driven constellation optimization mechanism, which can calculate feasible solutions for SPLD in polynomial time. Finally, we conduct extensive trace-driven simulations to verify \name's cost-effectiveness in constructing survivable and performant LSNs on demand, and showcase how \name can help optimize the incremental deployment and long-term maintenance of future satellite Internet.

\end{abstract}
\section{Introduction}
\label{sec:introduction}

Thanks to innovative breakthroughs in today's rocket industry and satellite communication technologies, low-earth orbit~(LEO) broadband satellites are greatly expanding the Internet frontier in recent years, propelling us into a new era of satellite Internet. ``NewSpace'' rising stars such as SpaceX's Starlink, OneWeb, Telesat and Amazon's Project Kuiper have taken center stage in constructing large-scale LEO satellite networks~(LSN) upon \emph{mega-constellations} which will consist of hundreds to thousands of inter-connected satellites, promising global Internet coverage on an unprecedented scale.

Since most emerging LSNs upon mega-constellations are still in plan or under heavy development, making an appropriate \emph{network design}, which determines the required amount of LEO satellites and connectivity configurations of an LSN, should be very critical for satellite operators. However, although ``mega-constellation'' conceptually refers to a large number of satellites, \emph{``\textbf{from a network perspective}, how many LEO satellites \textbf{exactly} does an LSN need''}, still remains an important but unclear problem facing the upcoming era of satellite Internet.

We observed two important trends related to LSN design. On the one hand, leading players such as SpaceX and Amazon plan to deploy their mega-constellations with a significantly large number of LEO satellites. For example, the first phase of SpaceX's Starlink and Amazon's Kuiper Project will consist of about 4,408 and 3,236 LEO satellites respectively. In addition, SpaceX also expects to extend the constellation size of Starlink to 42,000 in its next deployment phase. The key network benefits of deploying \emph{\textbf{more}} satellites in an LSN mainly include:
\begin{itemize}[leftmargin=*]
	\item \textbf{Better survivability.} 
	Nodes and links in an LSN inevitably suffer from failures due to a series of complex factors such as solar storm~\cite{solar_storm_destory_satellites}, radiation interference and hardware malfunctions~\cite{perumal2021small}. Deploying more  satellites in an LSN can effectively improve the survivability of the network since it provides more redundant backup links and paths.
	\item \textbf{Better performance.} LSNs target at providing broadband network services for global terrestrial customers. Deploying more LEO satellites with multiple high-throughput spot beams~\cite{wang2023satellite} can enable broader service coverage as well as higher constellation-wide network capacity~\cite{DELPORTILLO2019123}.
\end{itemize}

On the other hand, despite the network benefits, some recent voices have also pointed out that the high density of mega-constellations can involve extra sustainable problems such as:
\begin{itemize}[leftmargin=*]
	\item \textbf{Higher costs.} Undoubtedly, more satellites in an LSN indicate higher costs for satellite deployment and maintenance. Therefore, besides well-funded giant companies like SpaceX and Amazon, some other organizations are exploring viable paths to wisely design constellations with much less number of satellites~(\eg IRIS$^2$~\cite{iris_constellation} and Sfera~\cite{sfera_constellation} constellation).
	\item \textbf{Governance issues}~(\eg space conjunction and debris). As the constellation density increases, spacecraft conjunction incidents could become more frequent. More satellites can also lead to an increased accumulation of space debris which is a significant concern for space agencies and satellite operators as it poses a big risk to operational LSNs~\cite{chou2022towards}.
\end{itemize}

The academia has many previous efforts related to the LSN design~(or constellation design) from different aspects. As this paper will introduce in detail in \S\ref{sec:preliminaries}, some works focused on designing a satellite network~(or a constellation) to optimize the communication coverage~\cite{beste1978design,mortari2011design}, delay~\cite{liu2015delay}, or network capacity~\cite{bhattacherjee2019network, deng2021ultra}. However, these LSN design approaches did not consider guaranteeing network survivability  under various potential fallible elements in space. Some other works systematically studied the survivable network design problem~\cite{lau2007survivable,gupta2009online,ene2014improved} in conventional \emph{static} terrestrial telecommunication networks, but it is difficult to directly apply them in emerging LSNs where the entire network backbone is highly \emph{dynamic}. More recently, some researchers propose multi-tier space network design~\cite{zhu2022delay,chou2022towards} which incorporates geostationary~(GEO) satellites to forward inter-LEO-satellite traffic and reduce the amount of required satellites. However, integrating GEO satellites may inevitably increase the network delay due to the higher orbit altitude, and GEO satellites typically have limited capacity to serve a significantly large number of broadband users. Collectively, we argue that the network community still lacks a systematical and effective approach to guide the design, deployment and maintenance of future LSNs {from multiple network aspects} including network survivability, performance and corresponding costs.

In this paper, we fill this gap by conducting a comprehensive study on the LSN design problem from a network perspective. Specifically, we explore the research problem: \emph{given the requirements on network survivability and performance, how should a satellite operator design the LSN to satisfy various requirements with the minimum number of satellites?} We carry out our study and make three contributions as follows.

First, we formulate the \emph{survivable and performant LSN design~(SPLD)} problem, which aims to find the minimum number of required satellites in an LSN, while simultaneously satisfying various survivability and performance requirements such as: sufficient amount of redundant paths, necessary link capacity, and acceptable end-to-end delay for any terrestrial communication pair.
Second, to solve the tricky SPLD problem efficiently and effectively, we propose \name, a requirement-driven LSN optimization mechanism which can design an LSN in polynomial time, satisfying the survivability and performance requirements.

As the third contribution of this paper, we verify the cost-effectiveness of \name on constructing survivable and performant LSNs on demands. 
Extensive evaluations driven by real-world constellation information and satellite trajectory demonstrate that \name can effectively optimize the number of required LEO satellites, without invalidating both survivability and performance requirements. We also showcase the value of \name in several crucial stages during the LSN deployment and operation, such as facilitating resilient constellation design, optimizing the partial LSN deployment and providing insights for constellation maintenance.  
\section{Background and Related Work}
\label{sec:preliminaries}


\subsection{Preliminaries for LEO Satellite Networks~(LSN)}
\label{subsec:mega_constellation_for_satellite_Internet}

\begin{figure}[t]
	\centering
	\includegraphics[width=0.95\linewidth]{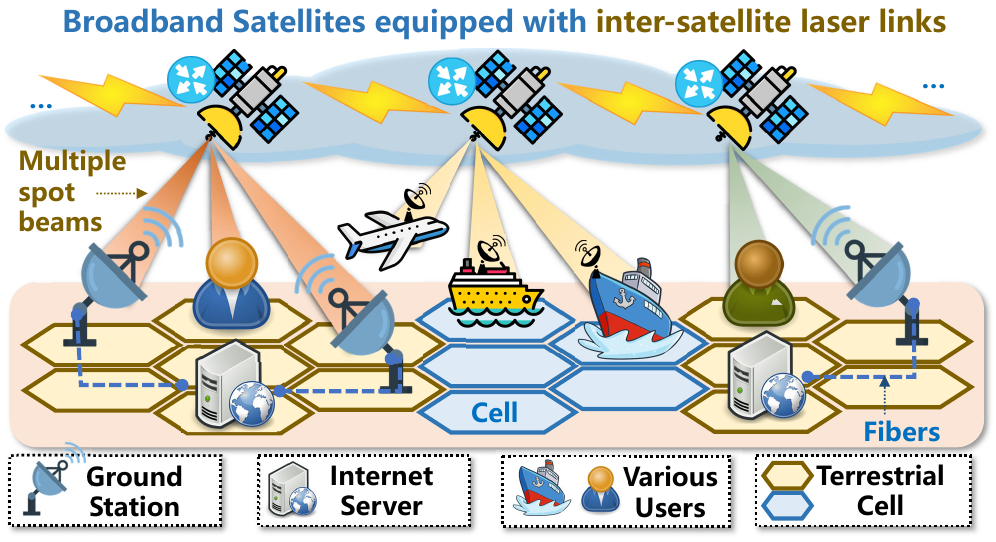}
	\caption{Typical architecture of LEO satellite networks~(LSN).}
	\vspace{-0.25in}
	\label{fig:lsn_architecture}
\end{figure} 

\noindent
\if
Recent ``NewSpace'' companies are actively deploying their mega-constellation with thousands of broadband satellites in low earth orbit (LEO).
\fi
\textbf{LSN architecture.} Figure~\ref{fig:lsn_architecture} plots a typical architecture of emerging LEO satellite networks~(LSN). Communication satellites can be equipped with high-speed inter-satellite links~(ISLs) and ground-satellite links~(GSLs) to construct an LSN, and then be integrated with terrestrial network infrastructures to provide ubiquitous, low-latency Internet services for various users.
The earth surface is divided into a large number of \emph{cells}, and emerging broadband satellites leverage multiple high-throughput spot beams to serve users spread in geo-distributed cells. In this architecture, LEO satellites perform two critical functions. First, these satellites work as \emph{space routers} to build an ``Internet backbone in space''~\cite{giuliari2020internet} and forward network traffic between any two cells especially for long-haul communications. Second, they also work as \emph{satellite ingresses} to provide last-mile network connectivity for terrestrial users and ground stations.

\noindent
\textbf{``Mega'' Internet constellations.} Unlike their predecessors~(\eg Iridium~\cite{irididum_constellation}), emerging satellite Internet constellations have grown dramatically in their density. SpaceX's Starlink is currently the most widely used commercial satellite Internet constellation. As of the date of June 2023, Starlink has launched more than 4000 LEO satellites and provided services for more than 1.5 million subscribers around the world~\cite{starlink_1_5_million}. 
Many other companies or organizations are also actively deploying their mega-constellations with thousands of LEO satellites, such as OneWeb~\cite{oneweb_constellation}, TeleSat~\cite{telesat_constellation}, and Amazon's Kuiper~\cite{kuiper_constellation}, with different constellation parameters~(\eg orbit altitude and inclination) but the same target~(\ie global Internet services).




\vspace{-0.05in}
\subsection{LSN Requirements: From A Network Design Perspective}
\label{subsec:LSN_requirement}

Since most LSNs are still under heavy development, making a proper LSN design in advance should be very crucial for satellite Internet operators. From a network design perspective, we generalize three important requirements for future LSNs.

\noindent
\textbf{Network survivability.} Typically, \emph{network survivability} refers to the ability of the network to maintain an acceptable level of service during various network failures. Note that unlike existing terrestrial networks where the core backbone is usually deployed in a protected environment, space routers in an LSN are operated in a failure-prone space environment, and thus are susceptible to various types of inferences or even malfunctions which are likely to cause network node or link failures. 
In addition, recent LEO broadband constellations are based on small satellites, which reduce their manufacturing time, but are relatively short-lived and prone to failure~\cite{lewis2017sensitivity,perumal2021small,solar_storm_destory_satellites}. Therefore, LSNs are expected to be survivable and resilient to failures and malfunctions in the complex space environment.

\noindent
\textbf{Low latency and high capacity.} Fundamentally, emerging LSNs are designed to provide Internet services for global users. For example, as plotted in Figure~\ref{fig:lsn_architecture}, a terrestrial user in remote or rural area can use the LSN to visit an Internet content server~(\eg a Web server), or communicate with other LSN users. Therefore, in addition to network survivability, LSNs are also expected to provide performance guarantee such as acceptable latency and sufficient capacity to accomplish good quality of experience~(QoE) for various Internet applications.

\noindent
\textbf{Appropriate scale.} The number of required satellites in a mega-constellation needs to be carefully considered since it affects the total cost of an LSN, which not only explicitly refers to the economic cost involved by the production and operation of a large number of satellites, but also implicitly indicates the orbit and radio spectrum occupation required by these communication satellites. 
In addition, recent works~\cite{chou2022towards,lawler2021visibility,huang2023co} have also pointed out that blindly increasing the number of satellites in a mega-constellation can incur a series of sustainable problems such as satellite conjunctions and space debris with serious security risks, light pollution, radio frequency interference and regulatory challenges.

The above network requirements suggest two diametrically opposed optimization directions for future LSN design. On the one hand, to improve the survivability and performance, an LSN should deploy \emph{more} satellites to enhance network capacity and build adequate redundant paths for any communication pair to facilitate fast network recovery in case of various failures. On the other hand, to reduce the cost and avoid serious sustainable issues, it is expected to reasonably deploy \emph{fewer} satellites. From a network perspective, how should an LSN operator judiciously design, deploy and maintain its LSN?

\subsection{Related Work}
\label{sec:related_work}

\noindent
Many previous works have investigated problems related to LSN design. We classify them into the following categories.


\noindent
\textbf{LSN~(constellation) design.} The satellite network and communication community has a long history studying on the constellation design for satellite communication systems. In the early stage, many constellation patterns were proposed~\cite{werner1995analysis,beste1978design,wang1993structural} to achieve global coverage, but they can only satisfy basic communication needs such as low-rate data communication for short message or voice services. More recently, various works have studied different aspects considered for satellite Internet mega-constellation deployments~\cite{jiang2018regional, liu2015delay,deng2021ultra,bhattacherjee2019network}. For example, Deng~\etal proposed an ultra-dense LEO constellation architecture, which minimizes the number of satellites while satisfying the backhaul requirement of each user terminal~\cite{deng2021ultra}. 
Motif~\cite{bhattacherjee2019network} is a new LSN topology design, exploiting repetitive patterns in the network to avoid expensive link changes over time, while still providing acceptable latencies and throughput. However, those approaches mostly focus on optimizing the coverage or quality of services~(QoS) of the LSN, but they ignore the joint requirements on network survivability and cost.



\noindent
\textbf{Survivable network design.} Over the past decades, the network community has a  number of works studying on the survivable network design problem~(SND) in conventional telecommunication networks. The original SDNP is to find a minimum cost subgraph satisfying various connectivity, hop, and performance requirements between network nodes. This leads to a wide variety of classical problems such as minimum cost flow and steiner tree \etc Notable efforts related to this area of research include Jain's 2-approximation algorithm for the edge-connectivity SDNP~\cite{jain2001factor}, together with its various variants with additional constraints~\cite{bansal2008additive,ene2014improved,orlowski2010sndlib,jain2001factor,gupta2009online,botton2013benders,xu2008survivable}. However, all those existing SDNP solutions are designed for conventional networks where network nodes and links are \emph{static}. They are difficult to be directly applied into emerging LSNs where the entire space backbone suffers from infrastructure-level frequent and endless dynamics.

\noindent
\textbf{Hybrid optimization. } There are some other recent works that propose to use a multi-tier, hybrid constellation design for satellite Internet~\cite{chou2022towards,zhu2022delay,cao2021multi,chen2022multi}. 
The core idea of these works is to use geostationary~(GEO) satellites with wider coverage to replace a portion of LEO satellites and thus reduce the total number of satellites. However, involving GEO satellites suffers from a fundamental limitation that, due to the higher orbit altitude and limited network capacity, it may result in higher user-perceived latency and reduced network scalability when serving a large amount of terrestrial customers. Besides, it also requires extra complicated negotiations and collaborations between different GEO and LEO satellite operators. Therefore, in this paper we focus on LSN design for a single operator.

\noindent
\textbf{Routing in LSNs.} Ultimately realizing high survivability and performance in an LSN requires the collaboration of both network design and routing techniques. In practice, to cope with various network failures, a survivable LSN design should provide sufficient redundant paths for high-value communication pairs. When a failure occurs, routing protocols should quickly detect the failure, update the routing table and forward traffic via other survival redundant paths to maintain the end-to-end reachability. Many previous efforts have studied efficient and resilient routing in LSNs~(\eg~\cite{pan2019opspf,zhang2021aser,lai2022spacertc,lai2023achieving}) and they complement our work in this paper.

Collectively, the limitations of existing network design efforts motivate us to discover a new cost-effective approach to guide satellite operators to design survivable and performant LSNs. 

\section{System Models and Problem Statement}
\label{sec:problem}

\subsection{System Models}
\label{subsec:system_models}

We start our quest by first introducing our system models based on the basic LSN architecture illustrated in Figure~\ref{fig:lsn_architecture}.

\noindent
\textbf{Constellation network model.} Let $\mathcal{S}=\{s_{1}, s_{2}, s_{3}, ...\}$ denote the set of all LEO satellites in an LSN. As described in \S\ref{sec:preliminaries}, these broadband satellites communicate with ground facilities~(\eg ground stations and user terminals) by radio GSLs. Recall that emerging satellite communication systems divide the earth surface into multiple \emph{cells}~(\eg by the well-know H3 method~\cite{uber_h3} in practice) and each cell can be covered by spot beams to build bi-directional communication links. Thus, we aggregate ground facilities into cells, and denote $\mathcal{C}=\{c_{1}, c_{2}, c_{3}, ...\}$ as the set of all cells served by the LSN.

LEO satellites are moving at a high velocity related to the earth. To model the impact of LEO dynamics on the network topology, we assume time is slotted, and denote a binary value ${I}^{t}_{ij}$ to indicate whether node $i$ and $j$ are visible to each other in slot $t$. Further, let binary value $e^{t}_{ij}=1$ indicate that there exists an active communication link between node $i$ and $j$ in slot $t$. Obliviously a communication link can be activated only if its two communication ends are visible to each other.   

Thus, an LSN can be formulated by a time-varying graph ${G}_{t} = ({V}, {E}_{t})$, where the vertex set ${V} = \mathcal{S} \cup \mathcal{C}$, and the edge set ${E}_{t}$ includes all active many-to-many GSLs and ISLs in slot $t$. The time-varying connectivity reflected by edge set $E_{t}$ characterizes the LEO dynamics in the LSN. 

\noindent
\textbf{Capacity model.} Emerging broadband satellites leverage multiple high-throughput spot beams to serve terrestrial users. These spot beams share the overall uplink/downlink capacity of the satellite~\cite{wang2023satellite}. We assume that one spot beam can serve one terrestrial cell. The uplink/downlink capacity of the spot beam between satellite ${i}$ and cell ${j}$ in slot $t$ is denoted as $Cap^{t}_{ji}$ and $Cap^{t}_{ij}$ (${i}\in \mathcal{S}$, ${j}\in \mathcal{C}$) respectively.
Let $Cap^{max}_{up}$ ($Cap^{max}_{down}$) denote the maximum uplink (downlink) capacity of a satellite, which is typically constrained by the power supplement of the satellite in practice. Then the real-time uplink (downlink) capacity of a certain satellite ${i}$ can be calculated as: $\sum_{j\in \mathcal{C}}{e^{t}_{ji} \cdot Cap^{t}_{ji}}$ ($\sum_{j\in \mathcal{C}}{e^{t}_{ij} \cdot Cap^{t}_{ij}}$). Similarly, the total achievable uplink~(downlink) capacity of a cell ${j}$ in a certain slot $t$ can be calculated as: $\sum_{i\in{\mathcal{S}}}{e^{t}_{ji}\cdot Cap^{t}_{ji}}$~($\sum_{i\in{\mathcal{S}}}{e^{t}_{ij}\cdot Cap^{t}_{ij}}$).
Although laser ISLs also have capacity limitations, according to recent references~\cite{DELPORTILLO2019123}, the capacity of laser ISLs is much higher than that of radio GSLs, and ISLs are unlikely to be the bottleneck in LSNs with existing space traffic steering~(\eg~\cite{lai2022spacertc}).

\noindent
\textbf{Communication demands.}
Let $\mathcal{D}=\{d_{1},d_{2},d_{3},...\}$ denote the set of all communication demands. Each $d_{k} \in \mathcal{D}$ is associated with a triplet $(\mathrm{src}(d_{k}), \mathrm{dst}(d_{k}), \mathrm{size}(d_{k}))$, where $\mathrm{src}(d_{k}), \mathrm{dst}(d_{k}) \in \mathcal{C}$ are the two communication ends~(\ie the source and destination cell) of $d_{k}$, and $\mathrm{size}(d_{k})$ is the traffic volume~(\eg datarate) of the demand. In practice, a concrete communication demand can be a video conferencing traffic or Web request/response carried by the LSN.

\noindent
\textbf{Survivability requirements.} For any demand $d$ associating with two distinct nodes $\mathrm{src}(d)=i$ and $\mathrm{dst}(d)=j$, an $[i\rightarrow j]$ path represents a sequence of nodes and edges in the LSN graph. There might be multiple available paths for a demand $d$, and we denote a collection of $[i\rightarrow j]$ as \emph{edge-disjoint} paths if no edge appears in more than one path. Further, we call $d$ associating with $i,j$ is \emph{R-edge connected}  if there are \emph{at least} R edge-disjoint paths between $i$ and $j$ in all time slots.

The number of edge-disjoint paths for a demand inherently indicates the LSN's ability to guarantee survivable communication for this demand.
Let $r_{ij}$ denote the \emph{survivability requirement} for demand $d$ associating with communication pair $(i,j)$, where $i$ and $j$ are two distinct terrestrial cells~($i,j\in\mathcal{C}$). $r_{ij}$ is assumed to be symmetric, \ie $r_{ij} = r_{ji}$. Specifically, requirement $r_{ij}$ indicates that it requires at least $r_{ij}$ edge-disjoint paths between $i$ and $j$ in all time slots. Intuitively, a higher $r_{ij}$ indicates more  redundant paths, and the communication session between $i$ and $j$ can resist against more link failures.

\vspace{-0.05in}
\subsection{Basic Survivable and Performant LSN Design~(SPLD)}
\label{subsec:basic_problem_formulation}


With all the system models defined above, we define the \emph{survivable and performant LSN design ~(SPLD)} problem: given an original LSN upon a mega-constellation, together with a series of survivability and performance requirements, how to appropriately optimize the network structure and find a sub-graph satisfying all requirements with the minimum amount of satellites? Specifically, we define $x(i)$ as a binary variable indicating whether a satellite $i\in\mathcal{S}$ is included or not in the sub-graph, and $T$ refers to the set of all time slots. Then the SPLD problem can be formulated as follows.

\noindent
\textbf{Objective:} \textbf{min} $\sum_{i\in \mathcal{S}} x(i)$,

\noindent
\textbf{Subject to:} 
\begin{equation}
	\label{eq:visibility_constraint}
	I^{t}_{ij} \geq e^{t}_{ij}, \forall i, j \in \mathcal{S} \cup \mathcal{C}, i \neq j, \forall t \in T,
\end{equation}
\begin{equation}
	\label{eq:selection_constraint}
	x(i) \cdot x(j) \geq e^{t}_{ij}, \forall i, j \in \mathcal{S} \cup \mathcal{C}, i \neq j,  \forall t \in T,
\end{equation}
\begin{equation}
	\label{eq:transmitter_constraint}
	\sum_{j\in \mathcal{S}} e^{t}_{ij} \leq N_{ISL}, \forall i \in \mathcal{S}, i\neq j, \forall t\in T, 
\end{equation}
\begin{equation}
\label{eq:up_capacity_requirement}
\sum_{\forall d: \mathrm{src}(d)=j}{\mathrm{size}(d)} \leq \sum_{i\in{\mathcal{S}}}{e^{t}_{ji}\cdot Cap^{t}_{ji}}, \forall j\in \mathcal{C}, \forall t \in T,
\end{equation}
\begin{equation}
	\label{eq:down_capacity_requirement}
	\sum_{\forall d: \mathrm{dst}(d)=j}{\mathrm{size}(d)} \leq \sum_{i\in{\mathcal{S}}}{e^{t}_{ij}\cdot Cap^{t}_{ij}}, \forall j\in \mathcal{C}, \forall t \in T,
\end{equation}
\begin{equation}
	\label{eq:satelliten_uplink_capacity_requirement}
	\sum_{j\in \mathcal{C}}{e^{t}_{ji} \cdot Cap^{t}_{ji}} \leq Cap^{max}_{up}, \forall i \in \mathcal{S}, \forall t \in T,
\end{equation}
\begin{equation}
	\label{eq:satelliten_downlink_capacity_requirement}
	\sum_{j\in \mathcal{C}}{e^{t}_{ij} \cdot Cap^{t}_{ij}} \leq Cap^{max}_{down}, \forall i \in \mathcal{S}, \forall t \in T,
\end{equation}
\begin{equation}
	\label{eq:redundant_path_constraint}
	\sum_{(i, j) \in \sigma(\overline{V})} e^{t}_{ij} \geq \textbf{max}_{\forall p\in V - \overline{V}, \forall q\in \overline{V}} \enspace r_{p q}, \forall \overline{V}\subset V, \overline{V}\neq\varnothing.
\end{equation}
Constraint~(\ref{eq:visibility_constraint}) indicates that two communication ends can establish a link between them only if they are visible and connectable to each other. Constraint~(\ref{eq:selection_constraint}) ensures that communication links can only be established if the associated satellites are included in the sub-graph. Constraint~(\ref{eq:transmitter_constraint}) indicates that for each satellite the total amount of activated ISLs can not exceed the number of available transmitters~($N_{ISL}$). Constraint~(\ref{eq:up_capacity_requirement}) and (\ref{eq:down_capacity_requirement}) guarantee the capacity requirement in both uplink and downlink directions for different traffic demands associated with geo-distributed cells. Constraint~(\ref{eq:satelliten_uplink_capacity_requirement}) and (\ref{eq:satelliten_downlink_capacity_requirement}) describe the GSL capacity limitation on each satellite.

The survivability requirements for communication demands are guaranteed by constraint~(\ref{eq:redundant_path_constraint}). $\overline{V}$ is a subset of the vertex set $V$, and $\sigma(\overline{V})$ represents the set of edges connecting $\overline{V}$ from $V$. This constraint ensures that the value of a minimum cut separating $p$ and $q$ is at least $r_{pq}$, implying that there are at least $r_{pq}$ edge-disjoint paths between the pair $(p,q)$.

\subsection{Extending SPLD with Delay Constraints}
\label{subsec:extended_problem_formulation}

The above basic SPLD formulation guarantees $r_{ij}$-edge connected for communication demand associated with $i,j$. However, in a practice, survivable redundant paths are also expected to satisfy certain delay requirements, \ie the length of these $r_{ij}$ edge-disjoint paths should not exceed a certain threshold to guarantee acceptable end-to-end delay. Therefore, we extend the basic SPLD with delay~(length) constraints.

\begin{figure*}[t]
	\centering
	\subfloat[The original LSN graph with a communication pair ($\mathrm{src}, \mathrm{dst}$).]{
		\label{fig:formulation_original_graph}
		\begin{minipage}[t]{0.95\columnwidth}
			\centering
			\includegraphics[width=\columnwidth]{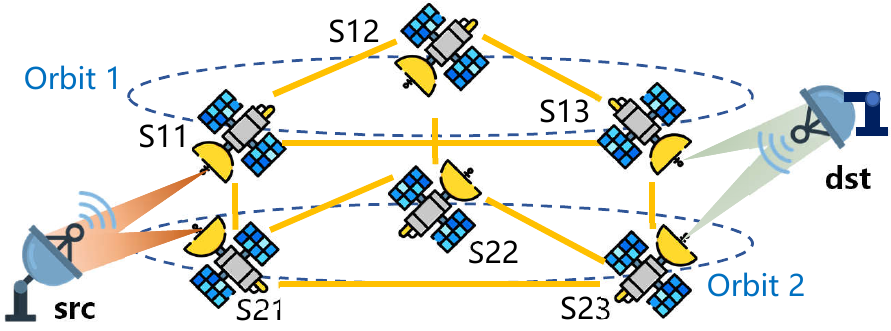}
		\end{minipage}
	}
	\subfloat[Convert the original graph to a directed layered graph representation.]{
		\label{fig:formulation_layered_graph}
		\begin{minipage}[t]{0.95\columnwidth}
			\centering
			\includegraphics[width=\columnwidth]{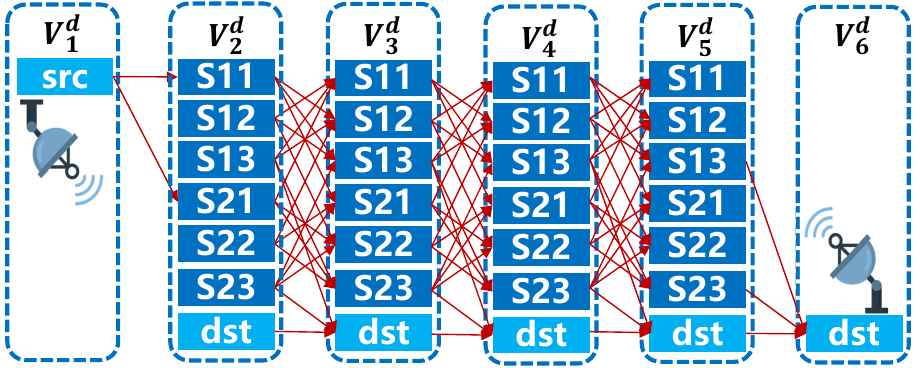}
		\end{minipage}
	}
	\vspace{-0.05in}
	\caption{The original network structure with two orbits and six satellites in total, and its layered presentation when $L=5$.}
	\vspace{-0.2in}
	\label{fig:graph_transform}
\end{figure*}

\noindent
\textbf{Graph transformation.} To formulate the delay-constrained SPLD, we convert the original non-directed $G_{t}$ to a directed layered graph. We then decompose the basic SPLD problem into $|\mathcal{D}|$ subproblems, one for each demand. 
Assume the path length of demand $d$ associated with $i$ and $j$ is expected to be lower than $L_{d}$~(or denoted by $L_{ij}$)
. Inspired by \cite{gvozdiev2018low}, the  value of $L_{d}$ can be set to   $ \lceil \lambda \cdot L^{sp}_{ij} \rceil$ where $L^{sp}_{ij}$ is the shortest path length between $i$ and $j$, and $\lambda \geq 1$ is a constraint factor. Further, we model each subproblem associated with demand $d$ by a directed graph composed of $L_{d}+1$ layers.


Based on the graph $G_{t} = (V, E_{t})$, we create a directed layered graph $G^{d}_{t} = (V^{d}, E^{d}_{t})$ for each $d$, where $V^{d} = V^{d}_{1} \cup V^{d}_{2} \cup ... \cup V^{d}_{L_{d}+1}$. We define $V^{d}_{1} = \{\mathrm{src}(d)\}$, $V^{d}_{L_{d}+1} = \{\mathrm{dst}(d)\}$, and $V^{d}_{l} = V - \{\mathrm{src}(d)\}$ for $l = 2, ..., L_{d}$. Let $v^{d}_{l}$ be the copy of $v \in V$ in the $l$-th layer of graph $G^{d}_{t}$. Then, the edge set of $G^{d}_{t}$ is defined as $E^{d}_{t} = \{(i^{d}_{l}, j^{d}_{l+1}) | (i,j)\in E_{t}, i^{d}_{l} \in V^{d}_{l}, j^{d}_{l+1} \in V^{d}_{l+1}, l\in{1,...,L_{d}}\} \cup \{(\mathrm{dst}(d)_{l}, \mathrm{dst}(d)_{l+1}) | l \in \{2, ..., L_{d}\}\}$. For simplicity, we define the directed edge between $i^{d}\in V^{d}_{l}$ and $j^{d}_{l+1} \in V^{d}_{l+1}$ by $(i,j,l)$, where  $d$ is omitted in the notation as it is often clear from the context. Each edge can carry at most one unit of flow and let $Cap^{d}_{t}(i,j)$ denote the link capacity of edge $(i,j)$ in $G^{d}_{t}$.

Assume a simplified LSN scenario as plotted in Figure~\ref{fig:formulation_original_graph}. This LSN contains two ground stations, two evenly spaced orbits and six satellites in total. Each satellite has three ISLs~(\ie $N_{ISL} =3 $). A communication demand associates with $\mathrm{src}$ and $\mathrm{dst}$. Each ground station has two visible ingress satellites at this time. Assume our goal is to find a minimal subgraph with at least 2 edge-disjoint paths for $(\mathrm{src}, \mathrm{dst})$~(\ie $r_{\mathrm{src},\mathrm{dst}}=2$), and the path length should not exceed five hops~(\ie $L_{\mathrm{src},\mathrm{dst}}=5$). Then we build a directed layered graph for the original undirected graph as plotted in Figure~\ref{fig:formulation_layered_graph}. For each node except for $\mathrm{src}$, we create a copy of the node from layer 2 to 5. For each edge $(i,j)$ in the original graph, we create a directed edge for their corresponding copies among different layers. By this transformation, \emph{we guarantee that any path from $\mathrm{src}$ to $\mathrm{dst}$ satisfies the $L$-hop constraint~(L=5)}.


Upon the transformed layered graph $G^{d}_{t} = (V^{d}, E^{d}_{t})$, we then formulate the delay-constrained SPLD problem as follows. 

\noindent
\textbf{Objective:} \textbf{min} $\sum_{i\in \mathcal{S}} x(i)$,

\noindent
\textbf{Subject to:}
\begin{align}
	\label{eq:flow_conservation}
	\sum_{\forall j:(j,i,l-1)\in E^{d}_{t}}{\omega^{(l-1)d}_{ji}} - \sum_{\forall j:(i,j,l)\in E^{d}_{t}}{\omega^{ld}_{ij}}  \nonumber \quad\quad \\ =
	\begin{cases}
		 -r_{d}, & if \quad i = \mathrm{src}(d)   \\
		 r_{d}, & if \quad i = \mathrm{dst}(d)_{l} \\
		 0, & otherwise
	\end{cases}
	, \quad\quad\quad\quad \\
	\nonumber \forall (i,j)\in E^{t}_{d}(i \neq j) , l\in\{1,...,L_{d}+1 \}, d\in \mathcal{D}, t\in T,
\end{align}
\begin{equation}
	\label{eq:edge_disjointness_constraint}
	\omega^{ld}_{ij} + \omega^{(l+1)d}_{ji}\leq 1, \forall (i,j)\in E_{t}, d\in \mathcal{D}, t\in T,
\end{equation}
\begin{equation}
	\label{eq:node_disjointness_constraint}
	\sum_{l\in\{2,...,L_{d}\}}{ \omega^{ld}_{ij} \leq x(i)}, \forall  (i,j)\in E_{t}, d\in \mathcal{D},  i\in \mathcal{S},t\in T,
\end{equation}
\begin{align}
	\label{eq:delay_constrained_capacity_requirement_downlink}
	\mathrm{size}(d) \leq \sum_{\forall j\in V^{d}}\omega^{L_{d}d}_{j,\mathrm{dst}(d)_{L+1}}\cdot Cap^{d}_{t}(j,\mathrm{dst}(d)_{L+1}), \mathrm{and}
	\\ \nonumber \mathrm{size}(d) \leq \sum_{\forall j\in V^{d}}\omega^{1d}_{\mathrm{src}(d),j}\cdot Cap^{d}_{t}(\mathrm{src}(d),j), \forall d\in \mathcal{D}, t\in T.
\end{align}

Each binary variable $\omega^{ld}_{ij}\in\{0,1\}$ describes whether the edge $(i,j,l)$ can carry flow for demand $d$ in the layered graph $G^{d}_{t}$. Note that $\omega^{ld}_{ij}=0$ if $l=0$. Let $r_{d}$ denote the survivability requirement for $d$. Constraint~(\ref{eq:flow_conservation}) is the flow conservation constraints at every node of the layered graph that guarantee that $r_{d}$ units of flows go from $\mathrm{src}(d)$ to $\mathrm{dst}(d)$, and it also ensures the $r_{d}$ survivability.  Constraint~(\ref{eq:edge_disjointness_constraint}) avoids local flow loops. Moreover, constraint~(\ref{eq:node_disjointness_constraint}) guarantees that edges associated with $i$ can be established to carry traffic only if $i$ is included in the sub-graph, and guarantees the edge-disjointness of the paths. Constraint~(\ref{eq:delay_constrained_capacity_requirement_downlink}) indicates the link capacity requirement for each demand.

The solution space of the integer programming formulation of our basic and delay-constrained SPLD problem is intractable for exhaustive search. Even if we set all $r$ to 1, the SPLD problem in a single time slot can be converted to the classic Steiner Tree Problem which is known to be NP-hard.
Our preliminary results show that the problem becomes intractable to solve even for moderately-sized instances with hundreds of satellites. Hence solving the SPLD problem requires the development of more efficient methods to obtain feasible solutions.

\section{Requirement-Driven LSN Optimization}
\label{sec:algorithm}

\subsection{\name Overview}
\label{subsec:approach_overview}


\name exploits a basic idea that: while it is difficult to directly solve the SPLD problem and obtain the optimal solution, it is doable to determine whether a given LSN is feasible to meet survivability and performance requirements in polynomial time. 
Specifically, \name starts with an initial constellation state, then repeatedly tunes the constellation structure as well as the number of satellites in multiple rounds of iterations, and searches the feasible LSN design with the minimum number of satellites. 

Specifically, \name exploits the following steps to find a near-optimal LSN design satisfying various  requirements.

\begin{figure}[t]
	\centering
	\includegraphics[width=0.98\linewidth]{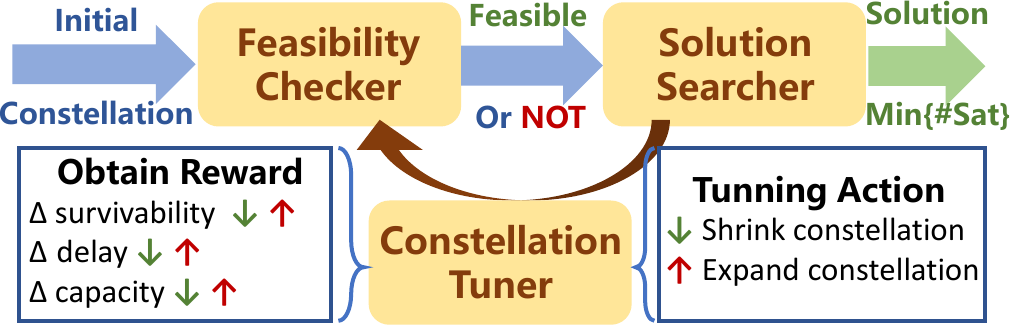}
	\vspace{-0.05in}
	\caption{High-level workflow of \name optimization.}
	\vspace{-0.25in}
	\label{fig:solution_overview}
\end{figure}

\begin{itemize}[leftmargin=*]
	\item \textbf{(1) Constellation initialization.} As illustrated in Figure~\ref{fig:solution_overview}, \name starts with an initial constellation pattern.
	This start point can be configured by a satellite operator, \eg using their current constellation design.
	\item \textbf{(2) Feasibility checking.} Based on the initial constellation, \name then exploits an \emph{efficient feasibility checking} process to examine whether the current state of the constellation can satisfy survivability and performance requirements. 
	\item \textbf{(3) Constellation tuning.} Based on the feasibility checking results, \name conducts \emph{constellation tuning} to update the LSN design. If the current constellation is a feasible one~(\ie can satisfy the survivability and performance requirements), we use a \texttt{Shrink} method to slightly reduce the amount of satellites. Otherwise, we invoke an \texttt{Expand} function to increase the density of the constellation.
	\item \textbf{(4) Solution search.}  Finally, \name iteratively repeats the feasibility checking and constellation tuning process. Among all feasible LSN designs, \name outputs the one with the minimal number of satellites.
\end{itemize}



\subsection{Searching LSN Design Solutions}
\label{subsec:constellation_searching}

At a high-level, \name searches the optimized LSN design as follows. First, it calculates the search range of the number of required satellites based on the orbital information. Second, it iteratively searches ``feasible points'' within this range. Finally, it chooses the feasible constellation configuration with the minimum number of satellites as the output decision.

\noindent
\textbf{Orbital characteristics.} In addition to the network constellation model described in \S\ref{subsec:system_models},  a satellite constellation can also be modeled by its orbital characteristics. For example, emerging Internet constellations~(\eg Starlink and Kuiper) follow the classic Walker Delta Constellation~\cite{walker1984satellite}, which contains a group of evenly-spaced orbits with the same inclination and altitude, and satellites in each orbit are also evenly spaced. This kind of constellations has an associated notation $[\mathrm{Inc}, \mathcal{O}, \mathcal{M}, \mathcal{H}]$, where $\mathrm{Inc}$ is the orbit inclination, $\mathcal{O}$ is the number of equally spaced planes, $\mathcal{M}$ is the total number of satellites in each orbit plane, and $\mathcal{H}$ is the relative spacing between satellites and adjacent planes. 
These orbital information can assist us to narrow down the search range of feasible solutions.

\noindent
\textbf{Determining the searching range.} \name searches the minimum number of required satellites in a range denoted as $[\mathcal{N}_{min}, \mathcal{N}_{max}]$. 
Initially, $\mathcal{N}_{max}$ is set to the number of satellites in the original constellation. We calculate the initial $\mathcal{N}_{min}$ by exploiting a key insight that: to guarantee at least $r_{ij}$ disjoint paths for communication pair between cell $i$ and $j$, there should be at least $r_{ij}$ visible satellites for $i$ and $j$. Hence, the initial $\mathcal{N}_{min}$ is calculated as the minimum number of satellites that ensures each cell $i$ has at least $\mathrm{max}\{r_{ij}\}, \forall j\in\mathcal{C}$ satellites in their transmission range during the service hours.


\noindent
\textbf{Searching process.} Algorithm~\ref{alg:consetllation_searching} illustrates the details of \name's searching process. Initially, \name invokes the \texttt{GetSurvivableBound} function to obtain the minimum number of satellites which describes the lower bound of the survivability requirement~(line 1-3). This guarantees that there are at least $r_{ij}$ visible satellites for cell $i$ and $j$. Iteratively, \name searches the feasible constellations within $[\mathcal{N}_{min},\mathcal{N}_{max}]$ by calling the \texttt{FeasibilityCheck} function. If a feasible constellation is found, then it calls \texttt{Shrink} to slightly reduce the constellation density. Otherwise, it calls \texttt{Expand} to increase the number of satellites~(line 4-14). Here we define a threshold $I_{limit}$ which constrains the maximum number of iterations. Finally, the feasible solution with the minimum number of satellites is selected as the result~(line 15). We next introduce the details of \texttt{FeasibilityCheck} and constellation tuning functions~(\texttt{Shrink} and \texttt{Expand}).

\setlength{\textfloatsep}{0.12cm}
\begin{algorithm}[t]
	\renewcommand{\algorithmicrequire}{\textbf{Input:}}
	\renewcommand{\algorithmicensure}{\textbf{Output:}}
	\caption{Searching Process of \name.}
	\label{alg:consetllation_searching}
	\begin{algorithmic}[1]
		\REQUIRE LSN graph $\mathcal{G}=\{G_{t}\}$, demand set $\mathcal{D}$, survivability requirement $\mathcal{R}=\{r_{ij}\}$, delay requirement $\mathcal{L}=\{L_{ij}\}$;
		\ENSURE A simplified LSN based on  $\{x_{i}\}$;
		\STATE $\mathcal{N}_{min} \leftarrow $\texttt{GetSurvivableBound}($\mathcal{G}, \{r_{ij}\}, \mathcal{D}$);
		\STATE $\mathcal{N}_{max} \leftarrow |V|$, $I\leftarrow 0$; /* \emph{iteration counter.} */
		\STATE $G^{result}_{list}\leftarrow \varnothing$, $[\mathcal{O}, \mathcal{M}, \mathcal{H}] \leftarrow $\texttt{GetConsInfo}($\mathcal{G}$);
		\STATE \textbf{while} $I\leq I_{limit}$ \textbf{do:} /* \emph{search the feasible solution.} */ 
		\STATE \quad /* \emph{call Algorithm~\ref{alg:feasibility_check} to check the current feasibility.} */
		\STATE \quad $\mathrm{feasible}\leftarrow$\texttt{FesibilityCheck}($\mathcal{G},\mathcal{D},\mathcal{R},\mathcal{L}$);
		\STATE \quad\textbf{if} $\mathrm{feasible}==True$ \textbf{do:}
		\STATE \qquad $\mathcal{N}_{max}\leftarrow \mathcal{G}.\mathcal{O} \cdot \mathcal{G}.\mathcal{M}$, $G^{result}_{list}$\texttt{.append}($\mathcal{G}$); 
		\STATE \qquad $\mathcal{G} \leftarrow $\texttt{Shrink}($\mathcal{G},\mathcal{N}_{min}$), $I\leftarrow I + 1$;
		\STATE \quad\textbf{else:}
		\STATE \qquad $\mathcal{N}_{min}\leftarrow \mathcal{G}.\mathcal{O} \cdot \mathcal{G}.\mathcal{M}$;
		\STATE \qquad $\mathcal{G} \leftarrow $\texttt{Expand}($\mathcal{G}$, $\mathcal{N}_{max}$), $I\leftarrow I + 1$;
		\STATE \quad\textbf{end if}
		\STATE \textbf{end while}
		\STATE \textbf{return} $\mathcal{G} \leftarrow \mathrm{argmin}_{\forall\mathcal{G}\in G^{result}_{list}}{(|\mathcal{G}.\mathcal{O} \cdot \mathcal{G}.\mathcal{M}|)}$.
	\end{algorithmic}
\end{algorithm}
\setlength{\floatsep}{0.12cm}

\vspace{-0.1in}
\subsection{Feasibility Checking}
\label{subsec:feasibility_check}

\begin{algorithm}[t]
	\renewcommand{\algorithmicrequire}{\textbf{Input:}}
	\renewcommand{\algorithmicensure}{\textbf{Output:}}
	\caption{\texttt{FeasibilityCheck} Process.}
	\label{alg:feasibility_check}
	\begin{algorithmic}[1]
		\REQUIRE LSN graph $\mathcal{G} = \{G_{t}\}$, demand set $\mathcal{D}$, survivability requirement $\mathcal{R}=\{r_{ij}\}$, delay requirement $\mathcal{L}=\{L_{ij}\}$;
		\ENSURE Feasibility decision $\mathcal{F}$;
		\STATE $\mathcal{F} \leftarrow True$; /* \emph{feasibility initialization.} */
		\STATE \textbf{for each} $t$ \textbf{in} $T$ \textbf{do:} /* \emph{check for each time slot.} */
		\STATE \quad\textbf{for each} $j$ \textbf{in} $\mathcal{C}$ \textbf{do:} /* \emph{initialize capacity.} */
		\STATE \qquad $AvaiCap^{up}_{j} \leftarrow \sum_{i\in{\mathcal{S}}}{e^{t}_{ji}\cdot Cap^{t}_{ji}}$;
		\STATE \qquad $AvaiCap^{down}_{j} \leftarrow \sum_{i\in{\mathcal{S}}}{e^{t}_{ij}\cdot Cap^{t}_{ij}}$;
		\STATE \quad\textbf{end for}
		\STATE \quad \textbf{for each} $d$ \textbf{in} $\mathcal{D}$ \textbf{do:}  /* \emph{check for each demand.} */
		\STATE \qquad /* \emph{Create a delay-constrained layered graph.} */
		\STATE \qquad $G^{d}_{t} \leftarrow $\texttt{GraphTransform}($d, G_{t}$)~\emph{/*  (\S\ref{subsec:extended_problem_formulation}). */};
		\STATE \qquad /* \emph{Check for survivability.} */
		\STATE \qquad $r\leftarrow$\texttt{CalculateMaxFlow}($\mathrm{src}(d), \mathrm{dst}(d), G^{d}_{t}$);
		\STATE \qquad \textbf{if} $r< r_{\mathrm{src}(d), \mathrm{dst}(d)}$, \textbf{return $\mathcal{F}\leftarrow False$}; 
		\STATE \qquad $s\leftarrow \mathrm{size}(d)$;/* \emph{Check communication capacity.} */
		\STATE \qquad \textbf{if} $s \leq AvaiCap^{up}_{\mathrm{src}(d)}$ \textbf{and} $s \leq AvaiCap^{down}_{\mathrm{dst}(d)}$ \textbf{do:}
		\STATE \qquad\quad $AvaiCap^{up}_{\mathrm{src}(d)}\leftarrow AvaiCap^{up}_{\mathrm{src}(d)} - s$;
		\STATE \qquad\quad $AvaiCap^{down}_{\mathrm{dst}(d)}\leftarrow AvaiCap^{down}_{\mathrm{dst}(d)} - s$;
		\STATE \qquad \textbf{else:} \textbf{return $\mathcal{F}\leftarrow False$}; \textbf{end if};
		\STATE \quad \textbf{end for} /* \emph{end of each demand check.} */
		\STATE \textbf{end for}	/* \emph{end of each time slot check.} */
		\STATE \textbf{return} $\mathcal{F}$. /* \emph{find a feasible constellation state.} */
	\end{algorithmic}
\end{algorithm}

Given an LSN described by a time-varying graph, the feasibility checking process determines whether the current LSN design can satisfy the survivability and performance requirements, as described in Algorithm~\ref{alg:feasibility_check}.
First, based on the constellation information, the feasibility checker calculates the available uplink/downlink capacity for each terrestrial cell~(line 3-6). Second, for each demand $d$, the checker transforms the constellation graph to an extended layered graph based on the methodology introduced in \S\ref{subsec:extended_problem_formulation}~(line 9). Specifically, assume that the source and destination cell of demand $d$ are $\mathrm{src}(d)$ and $\mathrm{dst}(d)$. Note that here we set the delay requirement for $d$ as $L_{\mathrm{src}(d),\mathrm{dst}(d)} = \lambda\cdot L^{SP}_{\mathrm{src}(d),\mathrm{dst}(d)}$, where $L^{SP}_{\mathrm{src}(d),\mathrm{dst}(d)}$ is the shortest path between $\mathrm{src}(d)$ and $\mathrm{dst}(d)$ in the LSN.

Further, the algorithm checks whether the current network can satisfy the survivability requirement~(line 11-12). Given the demand $d$ with the survivability requirement $r_{\mathrm{src}(d), \mathrm{dst}(d)}$, \texttt{CalculateMaxFlow} computes the maximum number of flows from $\mathrm{src}(d)$ to $\mathrm{dst}(d)$.
The LSN graph can meet the survivability requirement only if the maximum number of flows from $\mathrm{src}(d)$ to $\mathrm{dst}(d)$ is at least $r_{\mathrm{src}(d), \mathrm{dst}(d)}$. 
Finally, the feasibility checker examines whether the remaining uplink/downlink capacity for cell $\mathrm{src}(d)$ and $\mathrm{dst}(d)$ are sufficient to carry the traffic demand of $d$~(line 14-17), and outputs the feasibility result of the given constellation state.

\subsection{Constellation Tuning}
\label{subsec:constellation_tuning}

Based on the feasibility result, \name then performs operations to tune the constellation configuration. Intuitively \name's tuning process slightly shrinks the satellite density if a feasible configuration is found, or expands the constellation population in a reasonable way if its current form can not satisfy various requirements.

The design details of the \texttt{Shrink}() and \texttt{Expand}() functions are illustrated in Algorithm~\ref{alg:constellation_tuning}. Specifically, these functions perform binary search in $[\mathcal{N}_{min},\mathcal{N}_{max}]$. Note that to shrink~(expand) a constellation, one can reduce~(increase) the number of orbits or reduce~(increase) the number of satellites in each orbit. Here our tuning algorithms tune the constellation to make $\mathcal{O}$ and $\mathcal{M}$ closer~(line 3-7 and line 10-14). This principle comes from an important insight from the widely used Walker Delta constellations that \emph{the maximum number of hops of the shortest path between any two communication satellites is $\lceil (\mathcal{O} + \mathcal{M})/2 \rceil$, which obtains its minimum value when $\mathcal{O} = \mathcal{M}$}. In other words, since $\mathcal{N}_{min}$ guarantees the lower bound of survivability, making $\mathcal{O}$ and $\mathcal{M}$ closer can decrease the upper bound of delay between any communication pair in the LSN.

\begin{algorithm}[t]
	\renewcommand{\algorithmicrequire}{\textbf{\texttt{Expand:}}}
	\renewcommand{\algorithmicensure}{\textbf{\texttt{Shrink:}}}
	\caption{Constellation Tuning Algorithms.}
	\label{alg:constellation_tuning}
	\begin{algorithmic}[1]
		\REQUIRE (dynamic graph $\mathcal{G}$, upper bound $\mathcal{N}_{max}$)
		\STATE $\mathrm{current}\_{N} \leftarrow \mathcal{G}.\mathcal{O} \cdot \mathcal{G}.\mathcal{M}$;
		\STATE $\mathrm{target}\_{N} \leftarrow \lfloor \frac{(\mathrm{current}\_{N} + \mathcal{N}_{max})}{2} \rfloor$;
		\STATE \textbf{if} $\mathcal{G}.\mathcal{O} \leq \mathcal{G}.\mathcal{M}$ \textbf{do:}
		\STATE \ $\mathrm{add}$ $\lfloor \frac{(\mathrm{target}\_{N}-\mathrm{current}\_{N})}{\mathcal{G}.\mathcal{M}}\rfloor$ $\mathrm{orbits}$ $\mathrm{in}$ $\mathrm{total}$;
		\STATE \textbf{else:}
		\STATE \ $\mathrm{add}$ $\lfloor \frac{(\mathrm{target}\_{N}-\mathrm{current}\_{N})}{\mathcal{G}.\mathcal{O}} \rfloor$ $\mathrm{satellites}$ $\mathrm{per}$ $\mathrm{orbit}$;
		\STATE \textbf{end if}
		\ENSURE (dynamic graph $\mathcal{G}$, lower bound $\mathcal{N}_{min}$)
		\STATE $\mathrm{current}\_{N} \leftarrow \mathcal{G}.\mathcal{O} \cdot \mathcal{G}.\mathcal{M}$;
		\STATE $\mathrm{target}\_{N} \leftarrow \lfloor \frac{(\mathrm{current}\_{N} - \mathcal{N}_{min})}{2} \rfloor$;
		\STATE \textbf{if} $\mathcal{G}.\mathcal{O} \leq \mathcal{G}.\mathcal{M}$ \textbf{do:}
		\STATE \ $\mathrm{reduce}$ $\lfloor \frac{(\mathrm{current}\_{N}-\mathrm{target}\_{N})}{\mathcal{G}.\mathcal{O}} \rfloor$ $\mathrm{satellites}$ $\mathrm{per}$ $\mathrm{orbit}$;
		\STATE \textbf{else:}
		\STATE \ $\mathrm{reduce}$ $ \lfloor\frac{(\mathrm{current}\_{N}-\mathrm{target}\_{N})}{\mathcal{G}.\mathcal{M}}\rfloor$ $\mathrm{orbits}$ $\mathrm{in}$ $\mathrm{total}$;
		\STATE \textbf{end if}
	\end{algorithmic}
\end{algorithm}


\section{Evaluation}
\label{sec:evaluation}

\subsection{Experiment Setup} 
\label{subsec:simulation}

\noindent
\textbf{(1) LSN simulator.} 
Because most emerging LSNs are still under deployment or just provide limited accessibility, we evaluate \name by LSN simulation. To build a trace-driven simulation environment, we collect constellation information from the public regulatory documents~\cite{starlink_fcc,kuiper_fcc}, and extend StarPerf~\cite{lai2020starperf}, a state-of-the-art LSN simulator which can mimic the LEO dynamics and network behaviors of an LSN. Specifically,
we extend StarPerf with the ability of flexibly tuning the constellation structure as well as the ability for survivability assessment. We set the ground station distribution based on the public information provided by \cite{starlink_sx}. Further, we follow a recent study~\cite{giuliari2021icarus} to set the capacity of each laser ISL to 20Gbps, set the capacity of each shared GSL to 4Gbps, and set $N_{ISL}=4$ in our experiment.
For each experiment in this section, we simulate a complete regression period for the evaluated constellation.
Upon the LSN simulator, we implement our \name optimizer based on two open libraries: \texttt{Gurobi}~\cite{gurobi} and  \texttt{SkyField}~\cite{skyfield}, which is an astronomy package for high precision research-grade orbit analysis and trajectory calculation. 

\noindent
\textbf{(2) Traffic demand generation.} We combine Starlink's availability map~\cite{starlink_coverage_map} and a recent population-based traffic model used in \cite{bhattacherjee2019network} to generate the LSN traffic demand matrix in our experiments. Specifically, we generate the traffic demand for each terrestrial cell where  satellite service is ready based on the availability map, and the traffic volume is set proportional to its population size which can be obtained from \cite{world_population}. 

\vspace{-0.05in}
\subsection{Optimizing LSN Designs under Various Requirements}
\label{subsec:number_of_required_satellites}

We first verify \name's ability to optimize LSN designs under various requirements.
Specifically, we use \name to optimize two commercial mega-constellations Starlink and Kuiper. In particular, we use the first phase of Starlink~(4408 LEO satellites in 5 orbital shells with the altitudes between 540 km and 570 km) and Kuiper~(3236 satellites in total) as the initial constellation configuration.

\begin{figure}[t]
	\centering
	\subfloat[Starlink optimization.]{
		\label{fig:sat_survivability_optimization_starlink}
		\begin{minipage}[t]{0.44\columnwidth}
			\centering
			\includegraphics[width=\columnwidth]{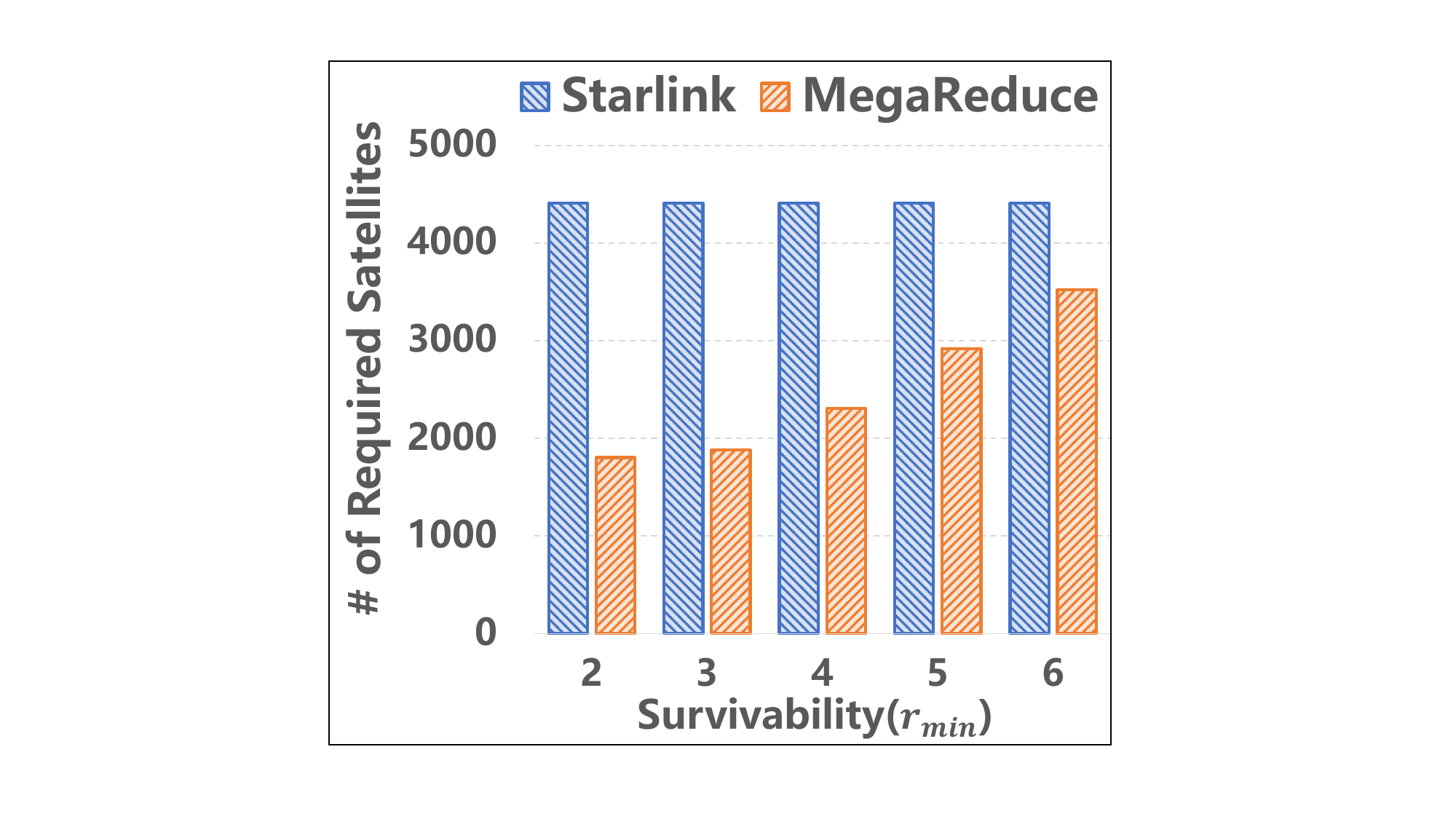}
		\end{minipage}
	}
	\subfloat[Kuiper optimization.]{
		\label{fig:sat_survivability_optimization_starlink}
		\begin{minipage}[t]{0.44\columnwidth}
			\centering
			\includegraphics[width=\columnwidth]{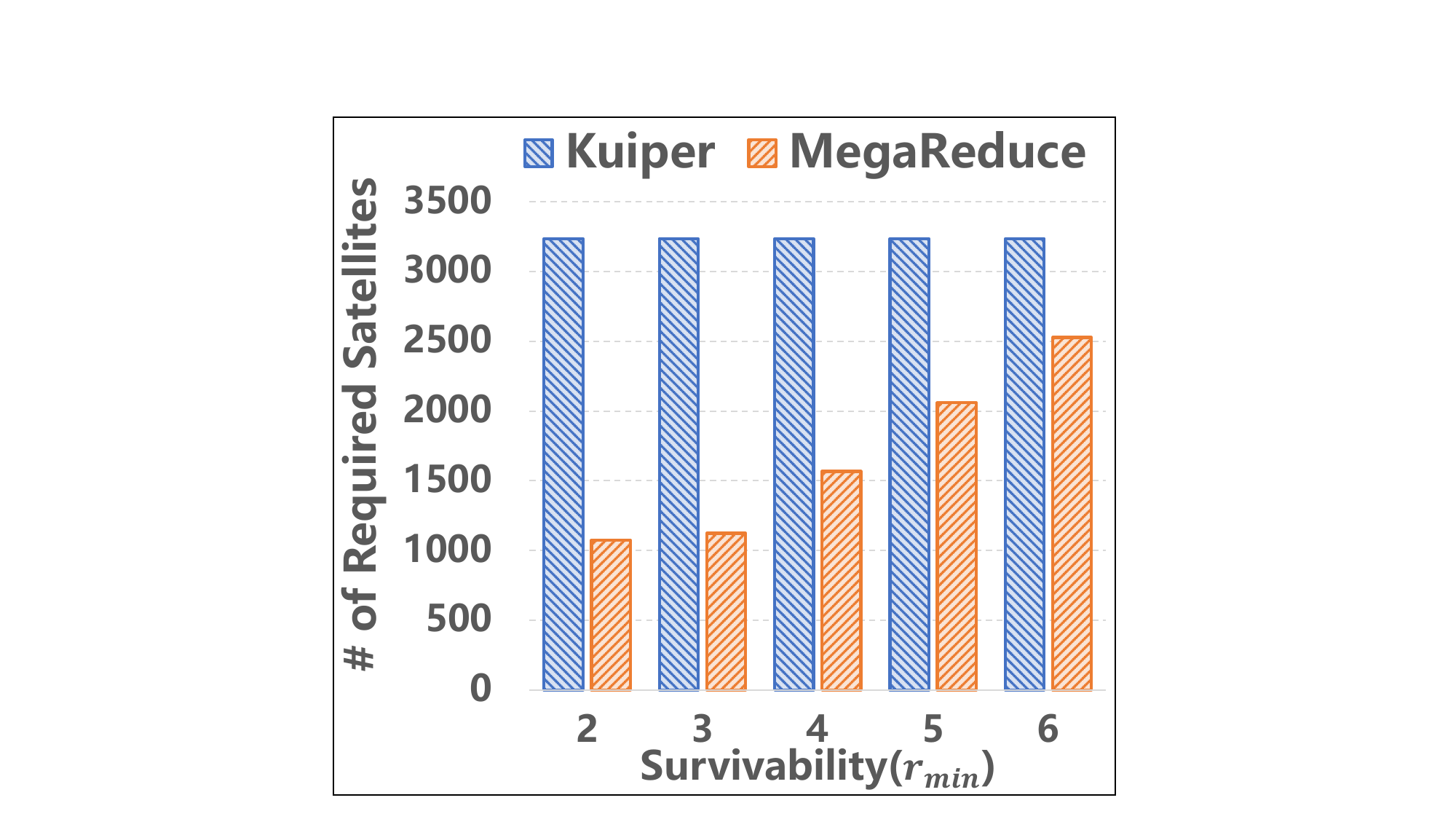}
		\end{minipage}
	}
	\vspace{-0.05in}
	\caption{Required \# of satellites v.s. survivability requirements.}
	\vspace{-0.2in}
	\label{fig:sat_survivability_optimization}
\end{figure}

\begin{figure}[t]
	\centering
	\subfloat[Starlink optimization.]{
		\label{fig:sat_capacity_optimization_starlink}
		\begin{minipage}[t]{0.44\columnwidth}
			\centering
			\includegraphics[width=\columnwidth]{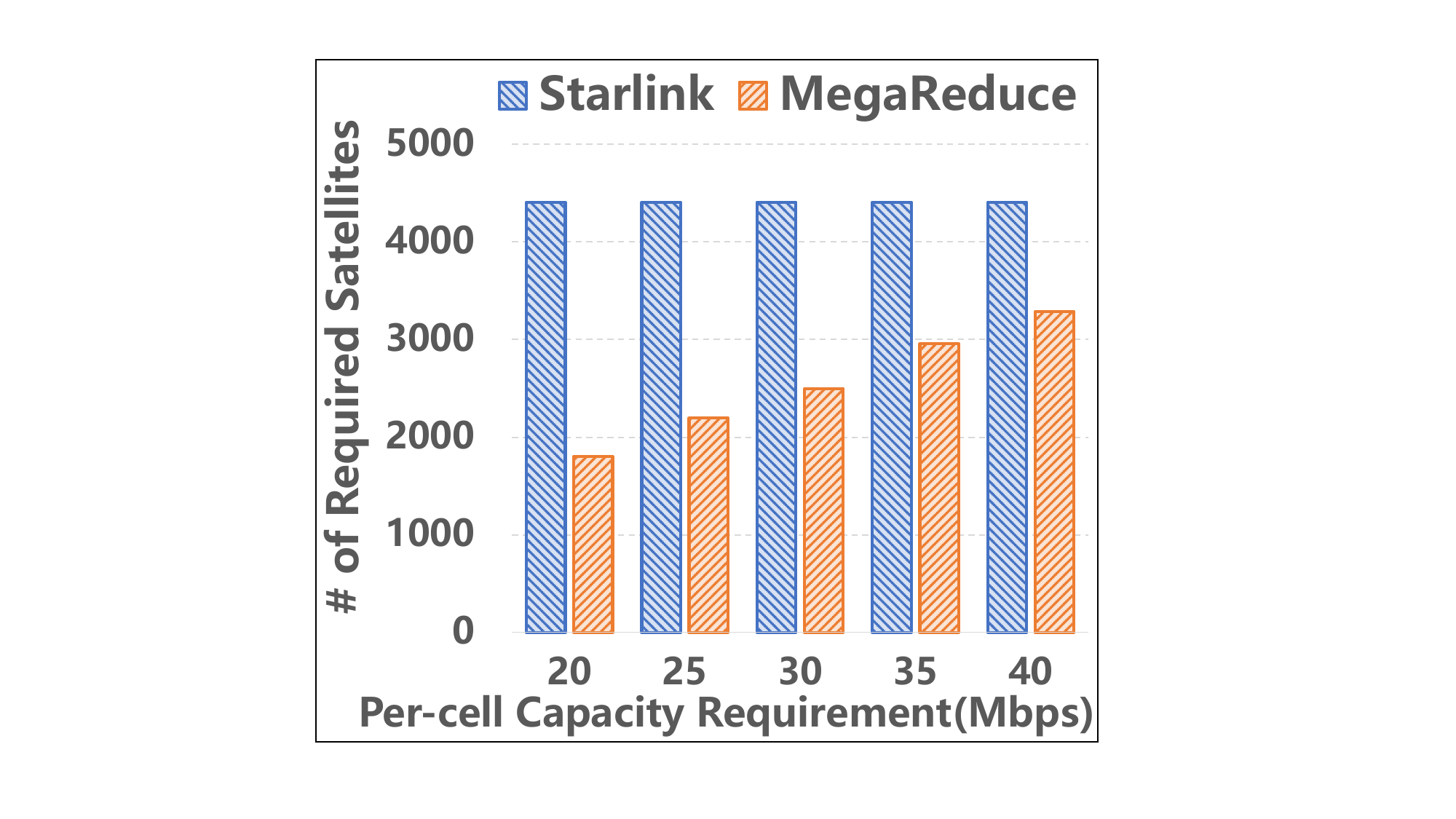}
		\end{minipage}
	}
	\subfloat[Kuiper optimization.]{
		\label{fig:sat_capacity_optimization_starlink}
		\begin{minipage}[t]{0.44\columnwidth}
			\centering
			\includegraphics[width=\columnwidth]{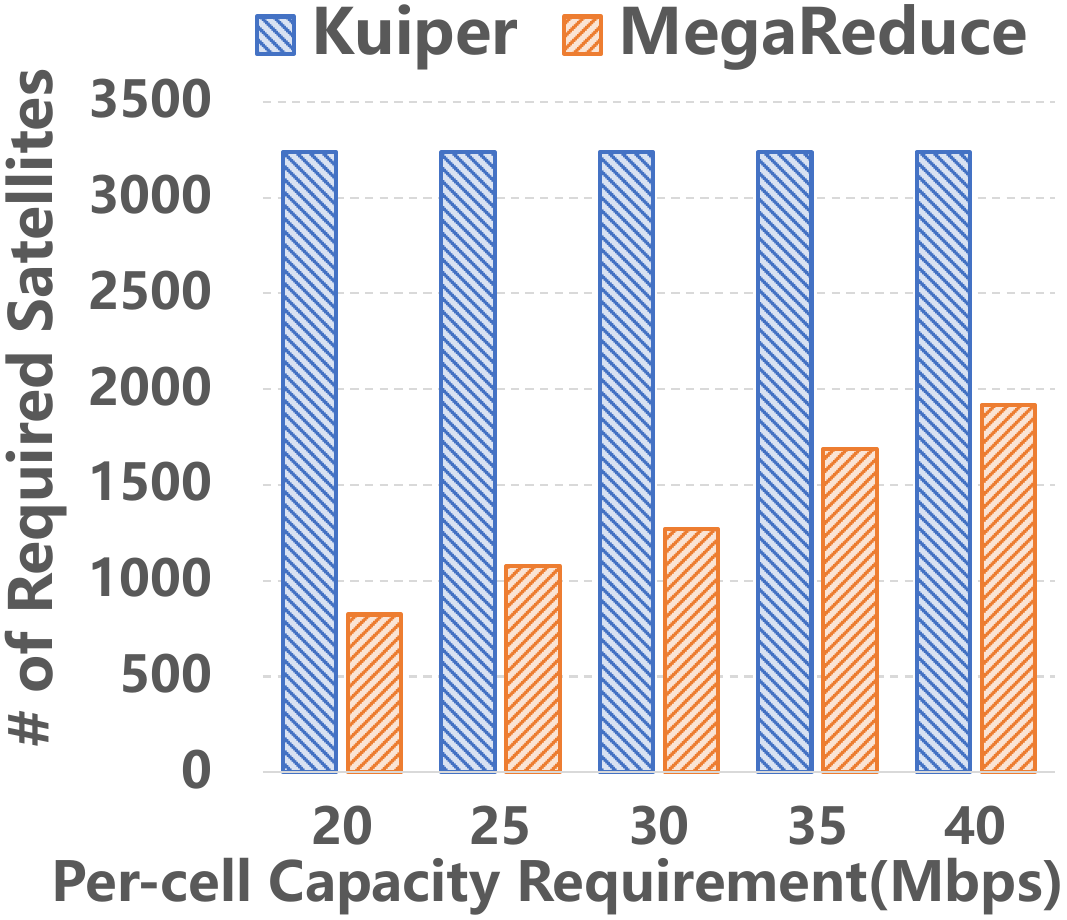}
		\end{minipage}
	}
	\vspace{-0.05in}
	\caption{Required \# of satellites v.s. capacity requirements.}
	\vspace{-0.05in}
	\label{fig:sat_capacity_optimization}
\end{figure}

\begin{figure}[t]
	\centering
	\subfloat[Starlink optimization.]{
		\label{fig:sat_delay_optimization_starlink}
		\begin{minipage}[t]{0.44\columnwidth}
			\centering
			\includegraphics[width=\columnwidth]{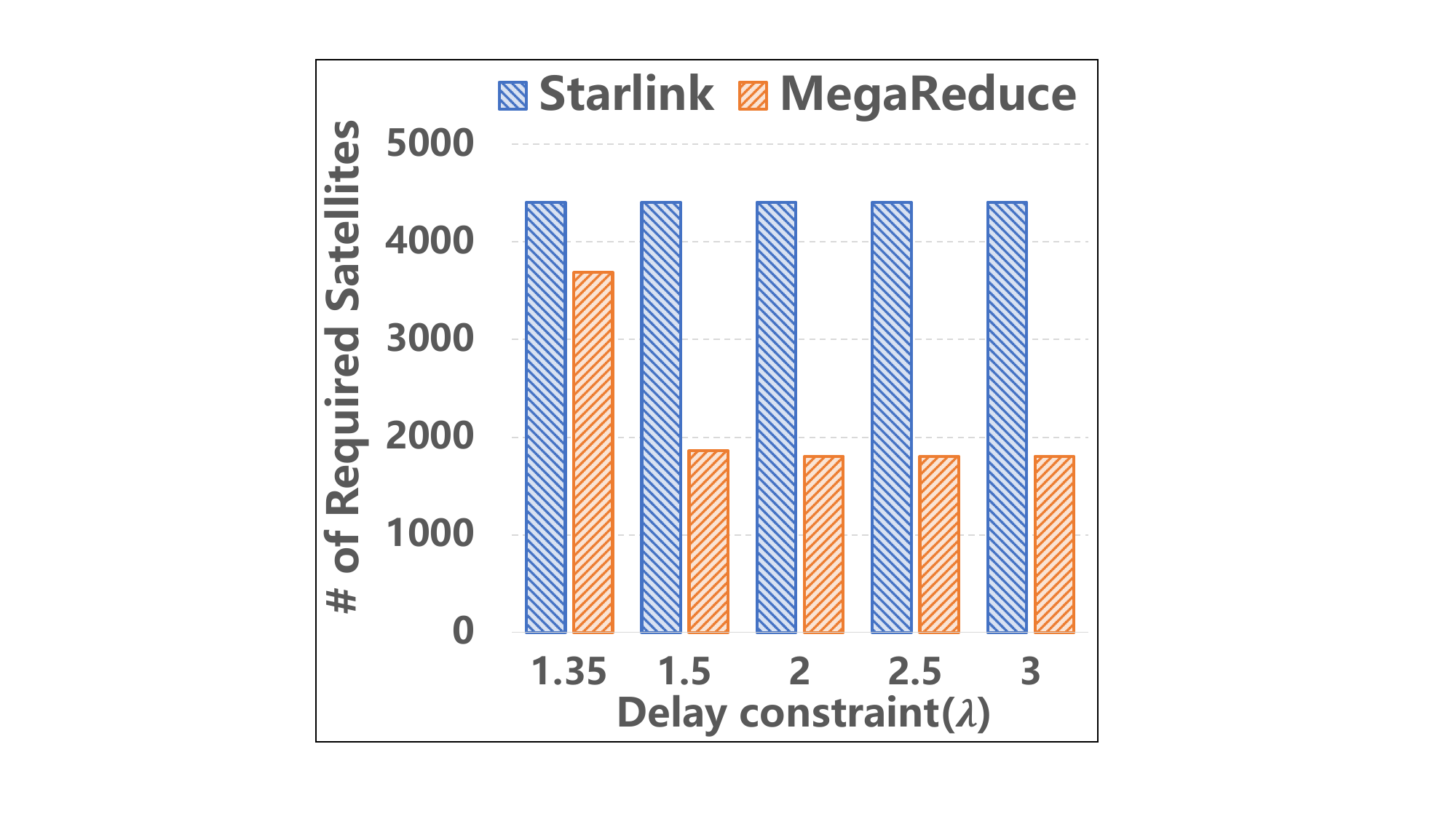}
		\end{minipage}
	}
	\subfloat[Kuiper optimization.]{
		\label{fig:sat_delay_optimization_starlink}
		\begin{minipage}[t]{0.44\columnwidth}
			\centering
			\includegraphics[width=\columnwidth]{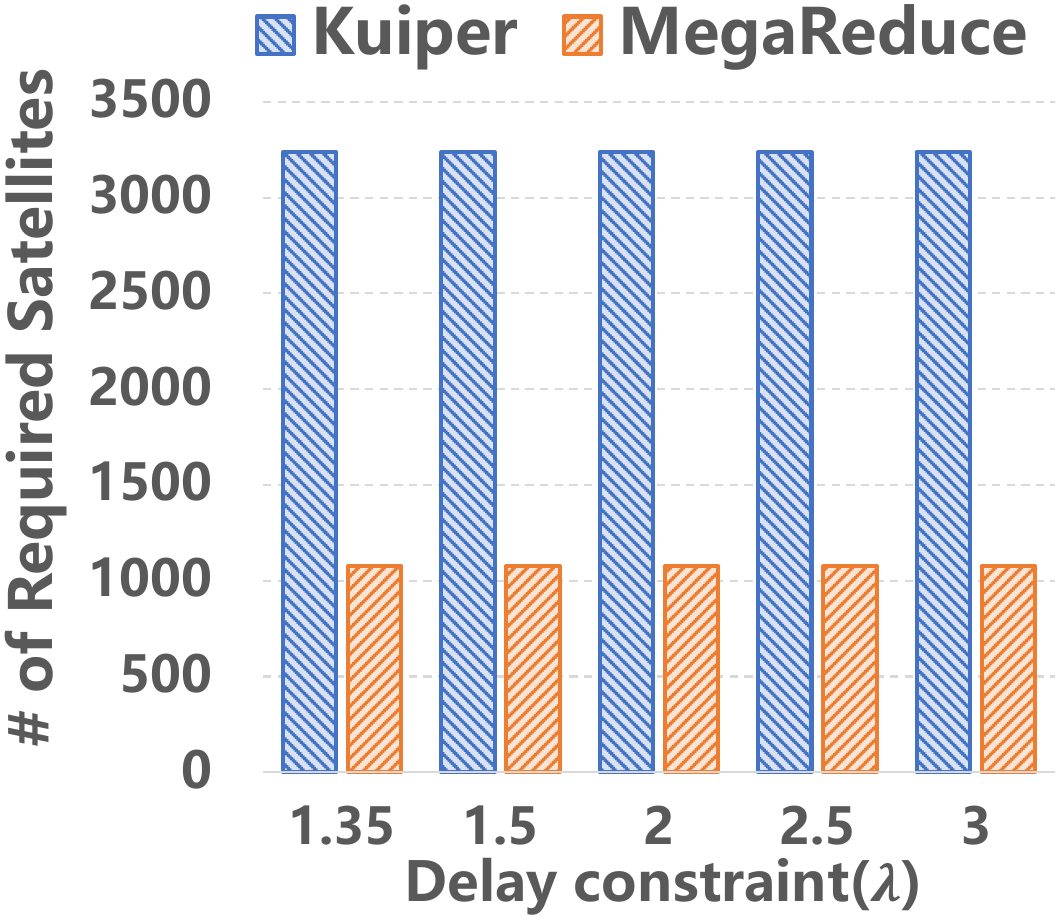}
		\end{minipage}
	}
	\vspace{-0.05in}
	\caption{Required \# of satellites v.s. delay constraints.}
	\vspace{-0.2in}
	\label{fig:sat_delay_optimization}
\end{figure}


Figure~\ref{fig:sat_survivability_optimization} plots \name's optimization on Starlink and Kuiper under various survivability requirements. 
Here the survivability parameter $r_{min}$ indicates that there should be at least $r_{min}$ edge-disjoint paths in the network for all communication pair in all time slots. The results in Figure~\ref{fig:sat_survivability_optimization} shows \name's ability to optimize the constellation size and reduce the number of required satellites while satisfying various survivability requirements. As the survivability increases, \name dynamically increases the constellation size and constructs more edge-disjoint redundant paths to improve the survivability of the LSN. 
Even in the case of $r_{min}=6$, \name can reduce 20.05\% and 21.88\% of the total number of required satellites for Starlink and Kuiper respectively.


Similarly, Figure~\ref{fig:sat_capacity_optimization} draws \name's optimization effectiveness under different capacity requirements. Specifically, we change the average per-cell capacity requirements form 20 to 40Mbps, and the results in Figure~\ref{fig:sat_capacity_optimization} demonstrate \name's ability to dynamically optimize the constellation size. The number of required satellites increases as the per-cell capacity demand increases, since more satellites can provide more high-throughput spot beams to serve terrestrial cells. 


In addition, we also verify \name's ability to adapt to various delay constraints, as shown in Figure~\ref{fig:sat_delay_optimization}. The delay constraint $\lambda$ indicates that the length of redundant paths for a communication pair $(i,j)$ should not exceed $\lceil \lambda \cdot L^{SP}_{ij} \rceil$, where $L^{SP}_{ij}$ is the length of the shortest path between $i$ and $j$. 
\name can dynamically adjust the number of required satellites to build redundant paths with similar lengths for communication pairs. As the value of $\lambda$ decreases, an LSN requires a denser constellation to provide more length-constrained redundant paths.

\subsection{Optimizing LSN Design under Various Orbital Parameters}
\label{subsec:impact_of_orbital_parameters}

\begin{figure}[t]
	\centering
	\subfloat[${r_{min}}=2$.]{
		\label{fig:sat_orbital_altitude}
		\begin{minipage}[t]{0.44\columnwidth}
			\centering
			\includegraphics[width=\columnwidth]{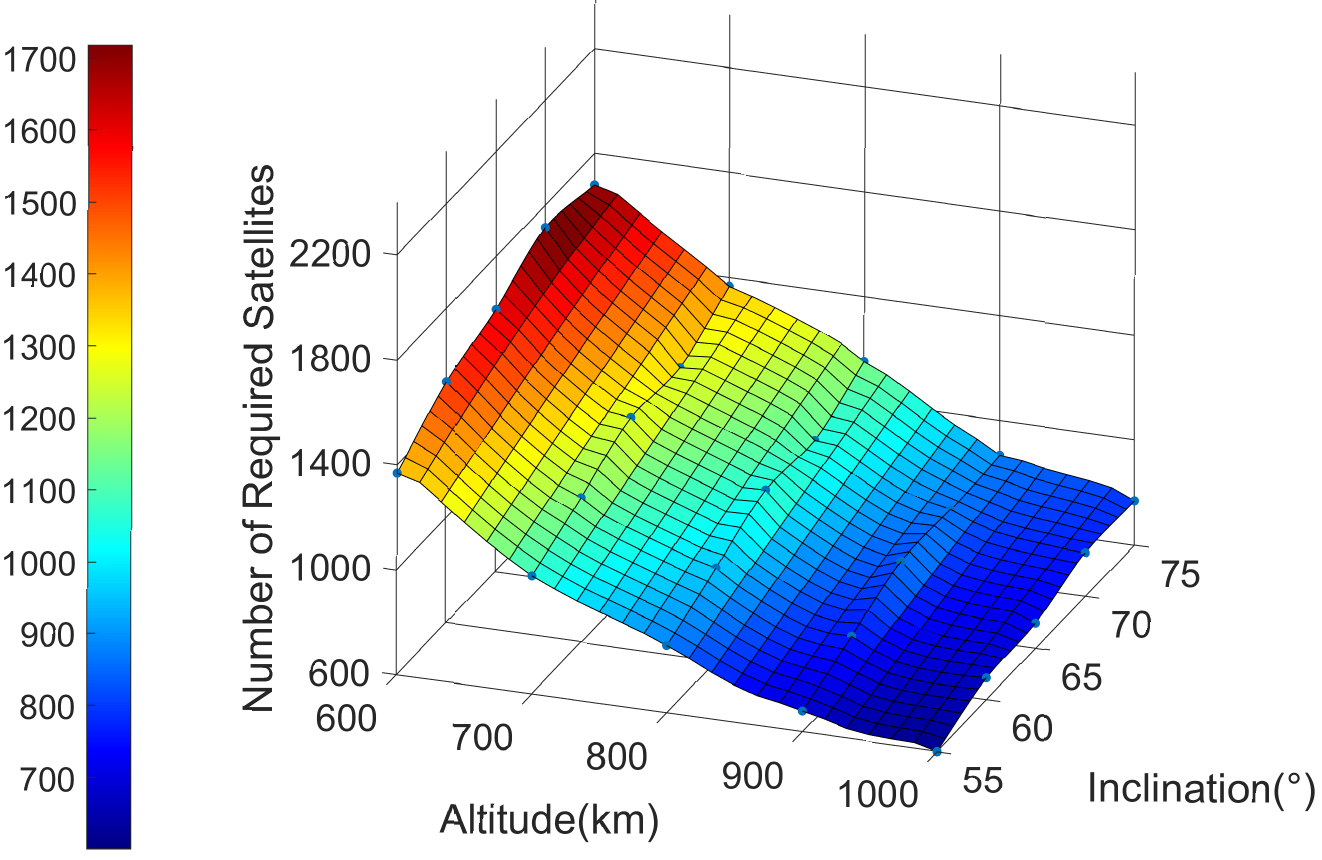}
		\end{minipage}
	}
	\subfloat[${r_{min}}=4$.]{
		\label{fig:sat_minimum_elevation_angle}
		\begin{minipage}[t]{0.44\columnwidth}
			\centering
			\includegraphics[width=\columnwidth]{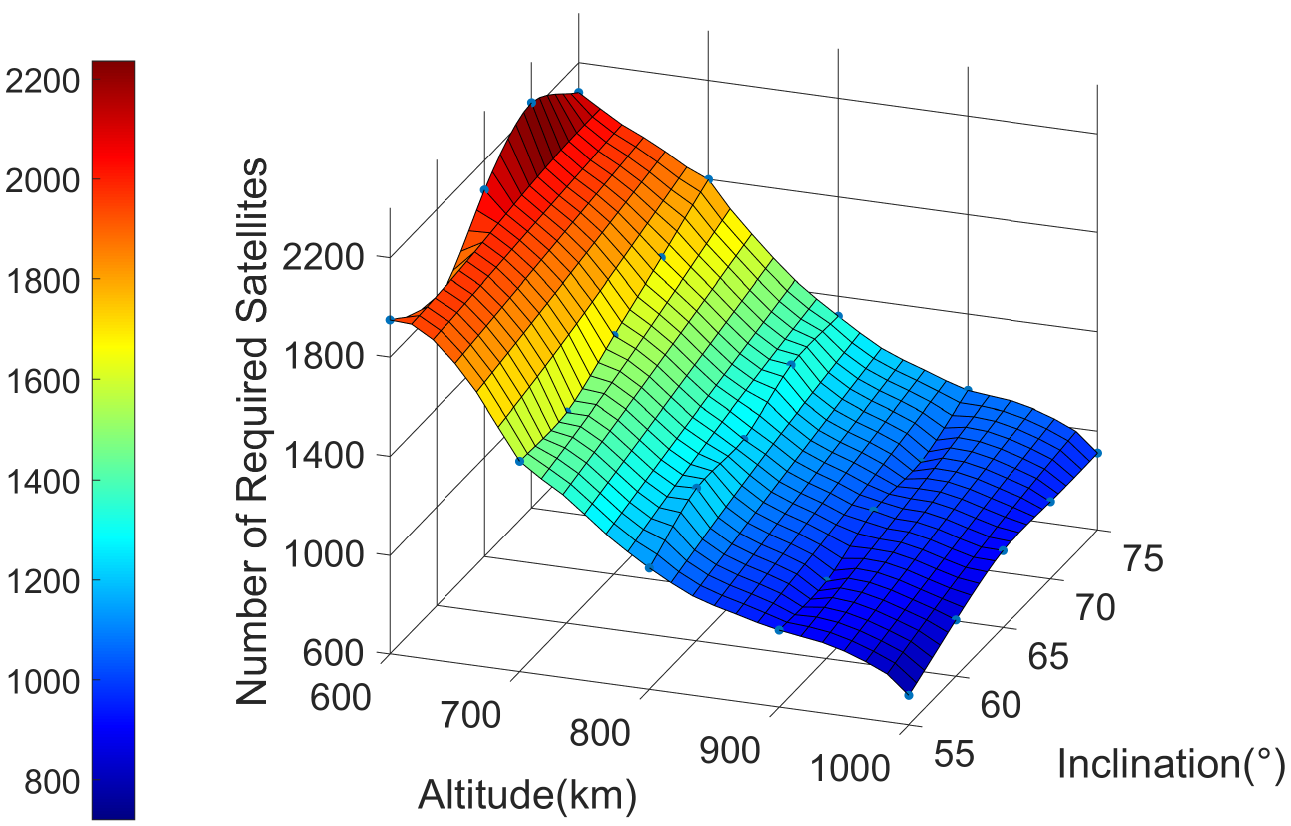}
		\end{minipage}
	}
	\vspace{-0.05in}
	\caption{Required \# of satellites under various orbital parameters.}
	\vspace{-0.05in}
	\label{fig:sat_orbital_analysis}
\end{figure}

Next we evaluate \name under various orbital parameters. Figure~\ref{fig:sat_orbital_analysis} shows a 3D surface plot of the required LEO satellite number versus the altitude of the satellites and the orbit inclination, under different survivability requirements~(${r_{min}}$). We make two main observations. First, the required number of satellites decreases as the orbit altitude increases. This is mainly because that a higher altitude indicates a wider satellite coverage, and thus one cell can be served by more satellites which increase the number of redundant links. However, the increased altitude can also involve higher space-to-ground propagation delay for LSN communications. Second, we find that the required number of satellites increases as the orbit inclination grows. This is because most communication cells are located inside $[-70^{\circ}, 70^{\circ}]$ latitude bands. A constellation with more inclined orbits covering this range can provide higher satellite density for those ``hot cells'' with more terrestrial users, and enable better network survivability as well as performance.

\subsection{Resilience Analysis}
\label{subsec:resilience_analysis}

\begin{figure}[t]
	\centering
	\vspace{-0.1in}
	\subfloat[Solar storm failure model.]{
		\label{fig:survivability_solar_storm_model}
		\begin{minipage}[t]{0.45\columnwidth}
			\centering
			\includegraphics[width=\columnwidth]{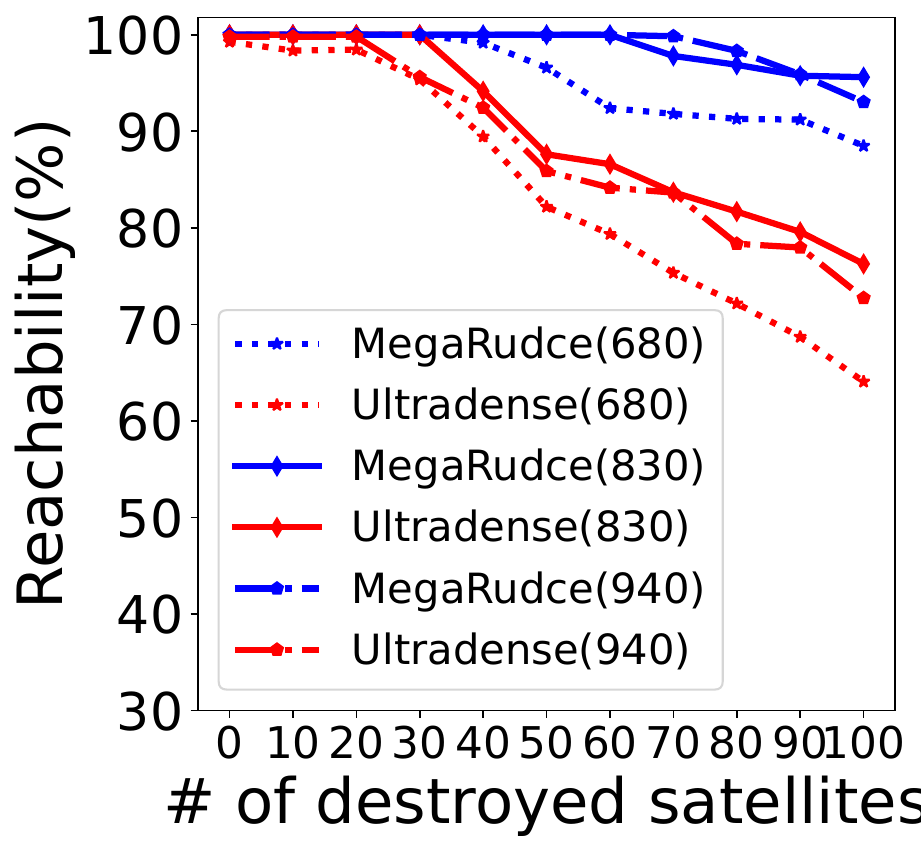}
		\end{minipage}
	}
	\subfloat[Random failure model.]{
		\label{fig:survivability_random_failure_model}
		\begin{minipage}[t]{0.45\columnwidth}
			\centering
			\includegraphics[width=\columnwidth]{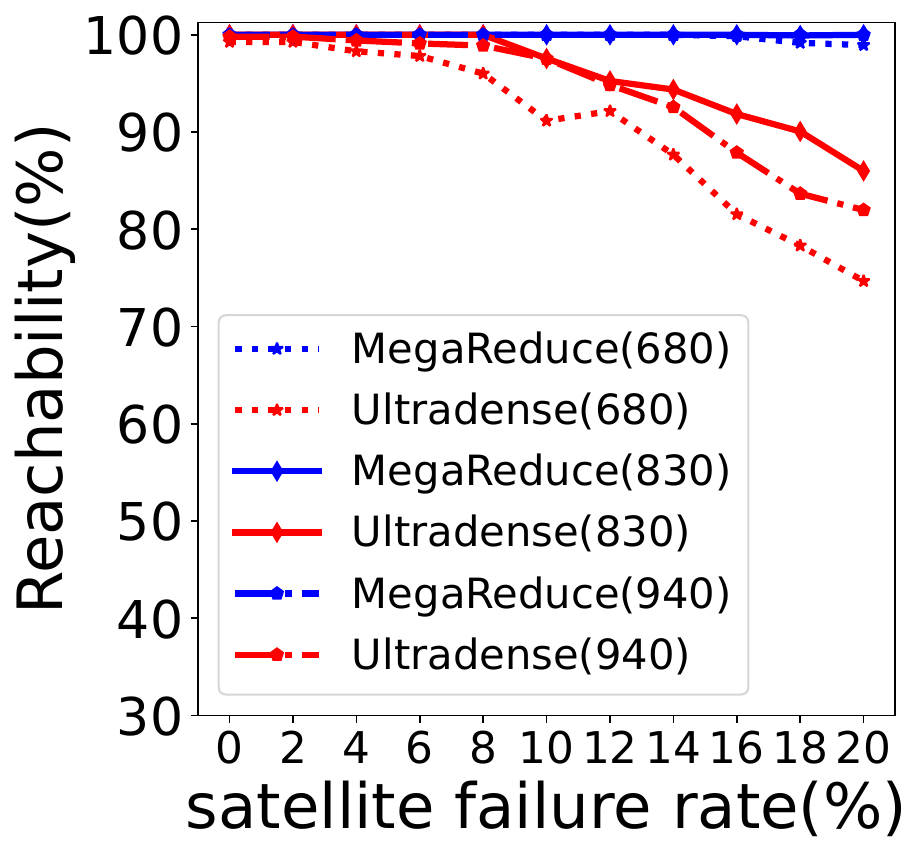}
		\end{minipage}
	}
	\vspace{-0.05in}
	\caption{Resilience analysis. The values in parentheses indicate the number of satellites used to build the LSN by each method.}
	\vspace{-0.2in}
	\label{fig:sat_resilience_analysis}
\end{figure}



Further, we evaluate the resilience of LSNs designed by \name.
We consider two failure models: (1) \emph{solar storm failure model}, in which a collection of nearby satellites are  destroyed at the same time~(\eg like the 49 Starlink satellites doomed on Feb. 4, 2022~\cite{solar_storm_destory_satellites}); and (2) \emph{random failure model}, in which satellites fail with a certain probability~(\eg due to hardware malfunctions or onboard component aging). We assume that
the high layer routing protocols can efficiently detect redundant paths and switch to the backup path if the current path fails. We compare \name with a recent constellation design approach called ultra-dense LSN design~\cite{deng2021ultra}. Figure~\ref{fig:sat_resilience_analysis} plots the reachability rate, which is calculated as the ratio of the number of inter-reachable pairs to the total number of communication pairs with different number of satellites used to construct the LSN. We observe that with the increasing number of destroyed satellites and failure rate, the total amount of reachable pairs reduces due to the lack of survival valid path. However, with the same constellation population, \name outperforms  UltraDense by achieving higher reachability ratio since it jointly optimizes the LSN design under survivability and performance constraints, guaranteeing a stronger ability to resist failures.



\subsection{Case Study}
\label{subsec:incremental_deployment}

\begin{figure}[t]
	\centering
	\subfloat[Incremental deployment.]{
		\label{fig:case_study_incremental_deployment}
		\begin{minipage}[t]{0.45\columnwidth}
			\centering
			\includegraphics[width=\columnwidth]{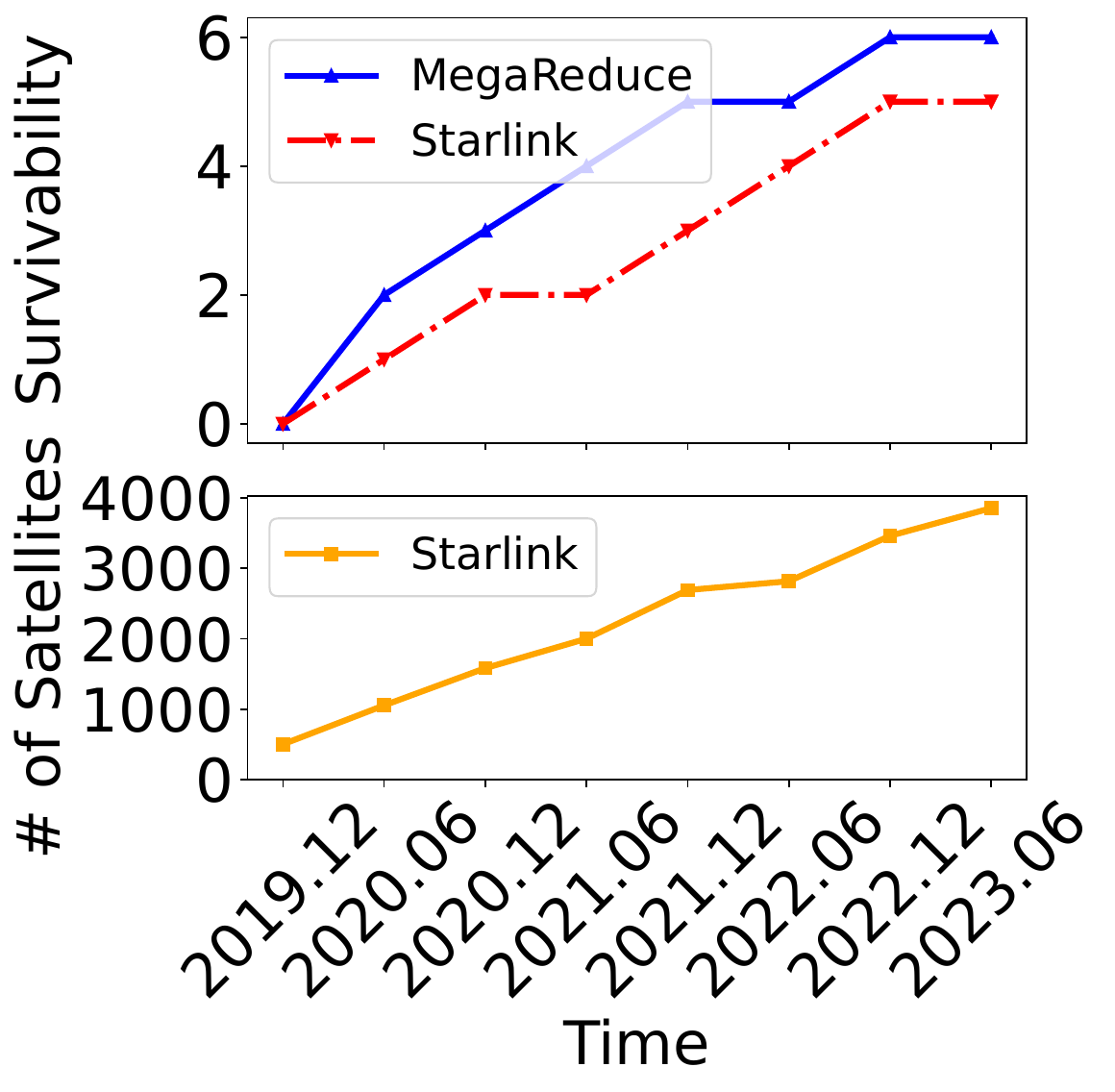}
		\end{minipage}
	}
	\subfloat[Constellation adjustment.]{
		\label{fig:case_study_constellation_compensation}
		\begin{minipage}[t]{0.45\columnwidth}
			\centering
			\includegraphics[width=\columnwidth]{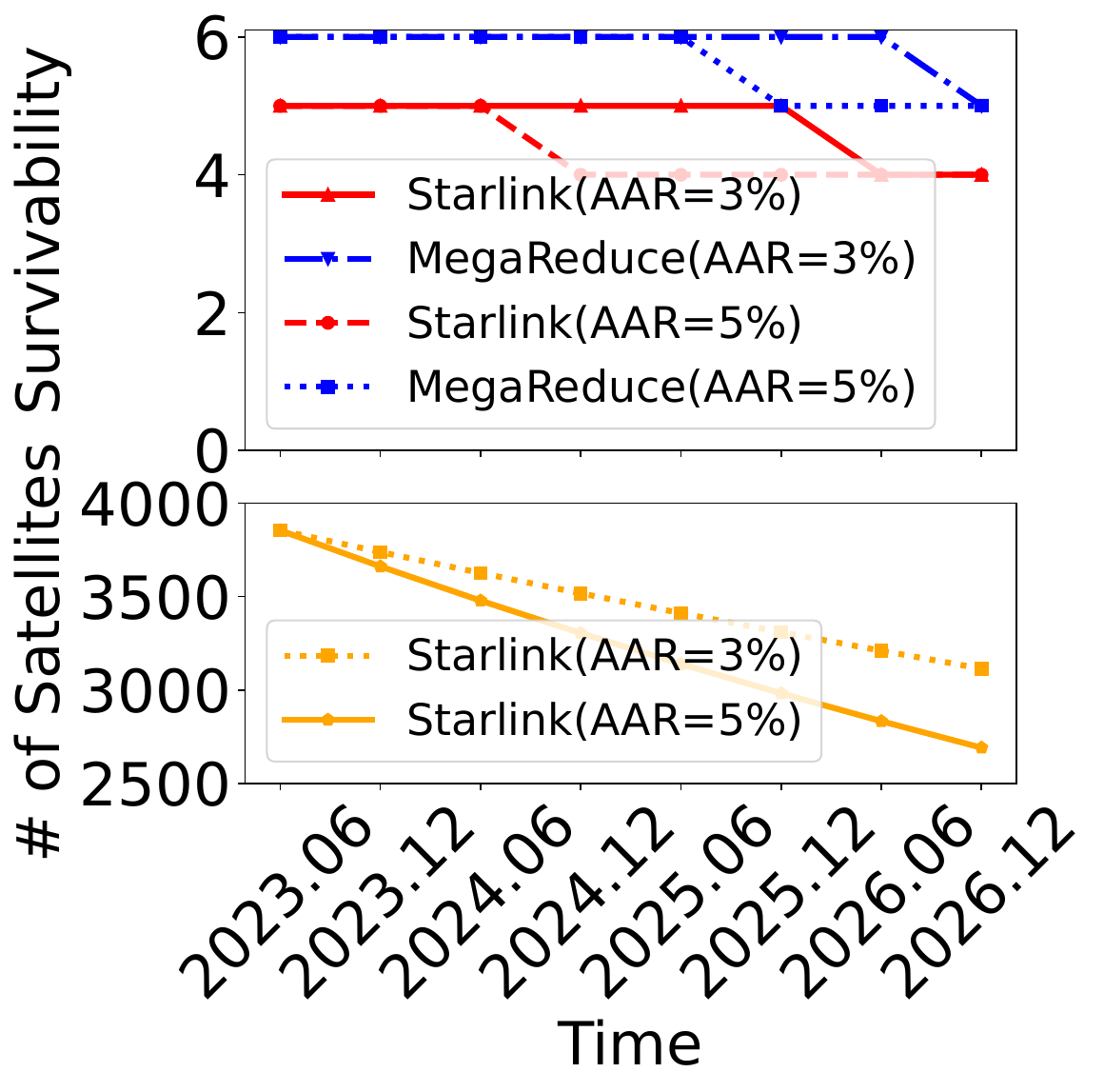}
		\end{minipage}
	}
	\vspace{-0.05in}
	\caption{\name can guide the incremental deployment and constellation adjustment scenarios for an LSN.}
	\vspace{-0.05in}
	\label{fig:case_study}
\end{figure}


Note that essentially \name represents a function $N^{sat}_{min}\leftarrow \mathcal{F}(req)$ where $N^{sat}_{min}$ stands for the minimum number of required satellites and $req$ refers to various requirements. With $\mathcal{F}$, we can easily calculate the inverse version of \name, \ie $req\leftarrow \mathcal{F}^{-1}(N^{sat})$, which indicates the achievable requirement under a given bound of satellite number. For example, if we have known that it needs at least 1500/1600 satellites to construct an LSN satisfying $r_{min}=5/6$, then we can deduce that given 1550 satellites, we can build an LSN with the survivability $r_{min}$ up to 5. Based on this insight, we perform two case studies to demonstrate how \name can help the deployment and maintenance of future LSNs. 

\noindent
\textbf{Incremental deployment.} Due to the high satellite density of emerging mega-constellations, typically satellite operators have to launch their satellites in multiple batches~(\ie incremental deployment) and it usually takes several years to completely deploy the entire LSN. As the LSN size gradually increases, it should be valuable if the LSN can maintain good survivability in its early stage even with only a few number of satellites deployed in orbit. We show that in each launch during the incremental deployment, \name can help an operator obtain a partial LSN design with the best survivability.

Figure~\ref{fig:case_study_incremental_deployment} plots the number of satellites and the survivability trend of real Starlink from Dec. 2019 to June. 2023. We obtain the historical data of Starlink from \texttt{SpaceTrack}~\cite{space_track}. With the deployment of Starlink, we observe that the survivability parameter $r_{min}$ increases from 0 to 5 within 3.5 years. Imaging in a parallel virtual world, we use the same number of satellites in each launch of real Starlink, but we follow \name to construct the LSN. Then we plot the survivability curve for two kinds of constellation design with the same number of satellites.
Results in Figure~\ref{fig:case_study_incremental_deployment} show that guided by \name, the constellation can achieve higher survivability even in the early stage of the incremental deployment.

\noindent
\textbf{In-orbit constellation adjustment.} Many mega-constellations are built upon small satellites
with shorter lifetime. Even after an LSN has been completely deployed, some satellites will gradually malfunction over time. \name can guide an operator to adjust the LSN topology and maintain a high survivability even though the LSN population decreases. Taking Starlink as an example. From the historical data~\cite{space_track} over the past 3 years, we find the annual decay rate~(AAR) of Starlink is about 2.6\%, which indicates that about 2.6\% satellites in the constellation become inactive per year. Figure~\ref{fig:case_study_constellation_compensation} plots the estimated size of Starlink with different AARs in the next 3 years. If no new satellites are launched, then the survivability will decrease gradually. We also plot the survivability curve if the operator adjusts the LSN structure in time when the number of satellites decreases following \name's solution, as plotted in Figure~\ref{fig:case_study_constellation_compensation}. \name can provide insights and guidelines for satellite operators to conduct in-orbit constellation adjustment to maintain high survivability when the mega-constellation gradually decays.

\vspace{-0.05in}
\section{Conclusion and Acknowledgement}
\label{sec:conclusion}
\vspace{-0.05in}

From a network perspective, how many satellites \emph{exactly} are needed to construct a survivable and performant LSN? In this paper, we first formulate the survivable and performant LSN design~(SPLD) problem, which aims at finding the minimum number of needed satellites while meeting various survivability and performance constraints. Then we propose \name, a requirement-driven, cost-effective LSN design approach which can calculate near-optimal solutions of SPLD in polynomial time. Finally, we conduct large-scale LSN simulation to verify the cost-effectiveness of \name for guiding the construction of survivable and performant LSNs.

We thank all anonymous reviewers for their valuable feedback. This work was supported by the National Key Research and Development Program of China (No. 2022YFB3105202) and National Natural Science Foundation of China~(NSFC No. 62372259 and No. 62132004). Hewu Li is the corresponding author. The authors have provided public access to their code and/or data at \url{https://github.com/SpaceNetLab/MegaReduce}.

\bibliographystyle{unsrt}
\bibliography{reference}

\end{document}